\documentclass[intlimits,twoside,a4paper]{article}

\usepackage{amsmath,amssymb}
\usepackage{graphicx}
\usepackage{graphics}
\usepackage{epstopdf}

\usepackage[T2A]{fontenc}
\usepackage[cp1251]{inputenc}

\usepackage{color}

\usepackage[eqsecnum]{cmpj2}
\issue{2015}{18}{4}{43002}
\doinumber{10.5488/CMP.18.43002}

\title[Coexistence of photonic and atomic Bose condensates]%
{Coexistence of photonic and atomic Bose-Einstein condensates in ideal atomic gases%
}
\author[N. Boichenko, Yu. Slyusarenko]{N. Boichenko\refaddr{label1},
       Yu. Slyusarenko\refaddr{label1, label2}}
\addresses{
\addr{label1} Akhiezer Institute for Theoretical Physics, NSC KIPT, 1 Akademichna St., 61108 Kharkiv, Ukraine
\addr{label2} Karazin Kharkiv  National University, 4 Svobody Sq., 61077 Kharkiv, Ukraine
}

\date{Received July 17, 2015}
\authorcopyright{N. Boichenko, Yu. Slyusarenko, 2015}

\begin{document}

\maketitle

\begin{abstract}
We have studied conditions of photon Bose-Einstein condensate formation that is in thermodynamic equilibrium with ideal gas of two-level Bose atoms below the degeneracy temperature. Equations describing thermodynamic equilibrium in the system were formulated; critical temperatures and densities of photonic and atomic gas subsystems were obtained analytically. Coexistence conditions of these photonic and atomic Bose-Einstein condensates were found. There was predicted the possibility of an abrupt type of photon condensation in the presence of Bose condensate of ground-state atoms: it was shown that the slightest decrease of the temperature could cause a significant gathering of photons in the condensate. This case could be treated as a simple model of the situation known as ``stopped light'' in cold atomic gas. We also showed how population inversion of atomic levels can be created by lowering the temperature. The latter situation looks promising for light accumulation in atomic vapor at very low temperatures.
\pacs 03.75.-b, 03.75.Hh, 03.75.Nt, 42.50.Ar
\end{abstract}

\section{Introduction}
\label{intro}

Bose-Einstein condensation is a vivid manifestation of quantum nature of macroscopic scale matter physics. This phenomenon is basic for many physical effects such as  superfluidity and superconductivity which have been known for a long time being used in practical applications. This is the fact that caused unabated interest to different Bose-Einstein condensate (BEC) related phenomena and effects. BEC direct experimental performance being obtained in alkali metal vapors (in this regard see~\cite{1,2,3}) opened up the prospect of experimental and theoretical predictions of new effects which are possible in systems with BEC~\cite{4,5}. For instance, such projections include the phenomenon of slow light~\cite{6,7,8} or even stopped light ~\cite{9} in a BEC, storage of light in atomic vapor at extremely low temperatures~\cite{10,11}. Studies~\cite{7,8} predicted the capability of controlling the group velocity of light in gases with BEC using an external magnetic field~\cite{12} as well as the possibility of using BEC for filtering optical electromagnetic signals~\cite{13}.  {Interesting effects ~\cite{Cumming,Heinzen}} and the effects associated with the passage of charged particles through systems with BEC were predicted in~\cite{14}.

To complete the overall academic research, experiments on BEC of photons were required. For several reasons there are few theoretical works devoted to this phenomenon (see, for example~\cite{15,16,17}). Indeed, when it comes to Bose-Einstein condensation, the possibility of such phenomena in gases as the simplest physical systems was studied first. As we have already mentioned, in first experiments the conditions of condensation were achieved within such systems  (see~\cite{1,2,3}). To get such a state in a bosonic gas it is required for particles to have a mass and to conserve their total number in the system. It is known that the mass of photon is zero in vacuum, and to observe the Bose condensation we need to reduce the temperature. It is difficult to find a method for lowering the temperature in a gas consisting only of photons and to create such a system is even more difficult. One may see that photons could behave like particles with non-zero mass, and to lower the temperature of that gas is possible when photons interact with the matter. However, in this case one should find the way how to compensate the loss of photons, because a decrease of the environmental temperature causes their absorption. All of these obstacles were recently overcome~\cite{18,19}: during simple and elegant experiments, BEC of free photons was obtained in dye-filled optical microcavity. Photons appeared in the system while pumping the dye solution using an external laser. Thermal equilibrium of the photon gas was reached as a result of absorption and re-emission. This made it possible to observe the condensation: due to the cut-off frequency, the effective photon mass became nonzero. Note that by cut-off frequency we mean the finite value of the wave vector of photons at zero frequency. Scientific community admitted this experiment to be a real breakthrough since it had been expected for a long time to receive a photonic condensate, and during the experiment it was observed at room temperature. This phenomenon may be also used in practical applications. For example, it could help gather and focus sunlight in solar panels at cloudy weather, create new sources of short-wavelength laser radiation, reduce the size of electronic microchips, etc.

It is obvious that since there are still a lot of issues to be discussed, the research needs to be continued in a number of ways. For example, is it possible to achieve BEC of photons in other systems? What temperatures are required? Is it possible to observe atomic and photonic Bose-Einstein condensates simultaneously? In this article the authors tried to answer the last question. We have studied the thermodynamic equilibrium of photonic and atomic gases at ultra-low temperatures, when BEC occurs in atomic gases. Note that the possibility of photons Bose-Einstein condensate formation in atomic nondegenerated gas was described in detail in~\cite{20}.

\section{Equations of thermodynamic equilibrium of photons and two-level ideal Bose-gas}
\label{general equations}

Actually, in this section we shall get even a more general task compared with the task posed in the Introduction. We shall study the possibility of Bose-Einstein condensation of photons that are in thermodynamic equilibrium with a two-level gas. Let us consider a gas below the degeneracy temperature that can consist of bosons as well of fermions.

As it was mentioned above, there were photons that are in thermodynamic equilibrium with two-level atom ideal gas at ultralow temperatures. This model implies that the atom has only two possible states, i.e., the ground state and the exited state. It means that when the atom absorbs a photon it becomes excited and, when the atom emits a photon, it changes the state from excited to the ground state. Thus, an excited atom can be considered as a bound state consisting of photon and non-excited atom. All these three components, i.e., photons, excited atoms and non-excited atoms, are in thermodynamic equilibrium. Let us assign subindex ``1''  to ground state physical characteristics, and subindex ``2''  to excited physical characteristics correspondingly (see also~\cite{20}). Distribution functions for two sorts of atoms~--- ``1''  and ``2''~--- are as follows:
\begin{equation}\begin{split}\begin{gathered}
 \label{1}
f_{\alpha _i}\left( \mathbf{p} \right)=\frac{1}{\exp \left[ \frac{{{\varepsilon }_{\alpha _i}}\left( \mathbf{p} \right)-{{\mu }_{i}}}{T} \right]\pm 1} \,, \\
{{\varepsilon }_{\alpha _i}}\left( \mathbf{p} \right)=\varepsilon _{\alpha _i}+\frac{{\mathbf{p}^2}}{2m}\,, \qquad
i=1,2\,.
\end{gathered}\end{split}
\end{equation}
Sign ``$+$''  in the function given above corresponds to the case of fermionic atoms, and sign ``$-$'' corresponds to the case of bosonic atoms. The quantum numbers are specified by parameters ${\alpha _1}$, ${\alpha _2}$ for all sorts of atoms. The values $\varepsilon _{\alpha _i}$ are the energy levels of stationary atoms; they are negative $\varepsilon _{\alpha_i}<0$, because an atom can be regarded as a bound state of some particles. Chemical potentials corresponding to both quantum-mechanical states are denoted as $\mu _i$, $i=1,2$. We imply that the atomic number in this system is conserved.

The photon distribution function looks as follows:
\begin{equation}
 \label{2}
{f_\textrm{ph}}\left( \mathbf{k} \right)=\frac{1}{\exp \left[ \frac{\hbar \omega \left( \mathbf{k} \right)-{{\mu }^*}}{T} \right]-1}\,,
\end{equation}
where $\omega \left( \mathbf{k} \right)$ is photon dispersion law, and $\mu ^*$ is photon chemical potential. The existence of non-zero chemical potential $\mu ^*$ points to the fact that the total number of photons $N_\textrm{ph}$ is conserved. It is important that the total number $N_\textrm{ph}$ of the photons should consist of free photons and of those photons absorbed by atoms (note that the number of excited atoms is equal to the number of absorbed photons). Current paper does not cover the reasons for the total number $N_\textrm{ph}$ conserving, but it could be supposed that some system of mirrors with high reflectivity provides that.  However, mirrors should be located far enough apart from each other to diminish the effect of boundary conditions so that they could be neglected.

From formulae (\ref{1}), (\ref{2}) we obtain these equations of the balance below:
\begin{equation}\begin{split}\begin{gathered}
 \label{3}
  N={g_{\alpha _1}}\sum\limits_{\mathbf{p}}{{f_{\alpha _1}}\left( \mathbf{p} \right)}+{g_{\alpha _2}}\sum\limits_{\mathbf{p}}{{f_{\alpha _2}}\left( \mathbf{p} \right)}, \\
 {N_\textrm{ph}}={g_\textrm{ph}}\sum\limits_{\mathbf{k}}{f_\textrm{ph}\left( \mathbf{k} \right)}+{g_{\alpha_2}}\sum\limits_{\mathbf{p}}{f_{\alpha _2}\left( \mathbf{p} \right)},
\end{gathered}\end{split}
\end{equation}
that follow from the fact of the number of atom $N$  and photon number  $N_\textrm{ph}$ conservation in the system. The parameter $g_{\alpha _1}$ (${g_{\alpha _2}}$) corresponds to degeneracy of atomic levels with the set of quantum numbers $\alpha _1$ ($ \alpha _2$). For instance, parameters $g_{\alpha _ 1}$,  $g_{\alpha _2}$ can take into account the spin state degeneracy, and $g^*$, i.e., the degeneracy of a photon with wave vector $\mathbf{k}$, may be caused by its polarization. Further we shall not take into account that characteristics of particles can depend on the spin.

To get a complete description of the system consisting of atoms and photons, we need to add the phase equilibrium condition to the system of equations
\eqref{3} (condition of chemical reaction in this regard see in \cite{21}, for example):
\begin{equation}
 \label{4}
\mu _1+\mu ^{*}=\mu _2\,.
\end{equation}
To calculate the sum over $\mathbf{k}$ in each equation in \eqref{3}, photon dispersion law needs to be specified. Subsequently the dispersion relation is assumed to be quadratic in the wave vector and it is given by the equation:
\begin{equation}
 \label{5}
\hbar \omega \left( \mathbf{k} \right)\equiv \hbar \omega \left( k \right)=\hbar \omega _0+\frac{p^2}{2m^{*}}\,,
\qquad
p\equiv \hbar k,
\end{equation}
where $\omega _0$ is the cut-off frequency of photon spectrum and $m^*$ is its effective mass. Let us remark that we used the quadratic photon dispersion relation and that it was done in \cite{18,19} with the difference that in these articles it had a two-dimensional wave vector that was caused by the microcavity parameters. Such a quadratic dependence on two-dimensional photon wave vector was also used in  \cite{26}  to obtain some  photonic gas features such as critical number of photons or spatial distribution of light luminosity in such a system. This paper does not explain the reasons why we use this exact  law and the dependence between the values $\omega _0$ and $m^*$. For example, the dispersion relation \eqref{5} can be obtained in case when formula
\begin{equation}
 \label{6}
\hbar \omega \left( k \right)=\hbar \sqrt{ \omega _{0}^{2}+  v^{2}   k^{2} }\,,
\end{equation}
gives photon energy for the photon in some medium. The expression \eqref{6} is similar to the expression for the energy of a relativistic object where $v$ is the phase velocity of light in a matter. If the wave vector $k$ satisfies the inequality $\left( v^{2} k^{2}/{\omega _0^2}\right)\ll 1$, formula \eqref{6} can be represented as follows:
\begin{equation}
 \label{7}
\hbar \omega \left( k \right)\approx \hbar {\omega _0}+\frac{{v^2}{{\left( \hbar k \right)}^2}}{2{{\omega }_0}\hbar }=\hbar {\omega _0}+\frac{{p^2}}{2{m^*}}\,,
\qquad p=\hbar k,
\end{equation}
and then it becomes possible to introduce the effective photon mass $m^*$ which can be addressed as photon ``rest energy'' in the matter, defined by:
\begin{equation}
 \label{8}
{m^{*}}=\frac{\hbar {\omega }_0}{v^2}\,.
\end{equation}
To denote the photon dispersion relation in certain matter and to get the cut-off frequency $\omega _0$ and speed $v$ of electromagnetic waves propagation in the matter one needs to formulate and to solve the dispersion relation of electromagnetic waves in the matter. Some basics of solving such problems in the case of electromagnetic waves propagation in ultracold atomic Bose gases can be found in \cite{22} and in \cite{7,8}. It is interesting that electromagnetic waves in plasma (see, e.g., \cite{23}) provide us with the dispersion laws that fully satisfy the formulae \eqref{6}--\eqref{8}. In particular, the dispersion relation for longitudinal waves in plasma is like formula \eqref{5}:
\begin{equation*}
{\omega_l}\left( \mathbf{k} \right)={\omega _0}\left[ 1+\frac{3}{2}{{\left( k{r_\textrm{D}} \right)}^2} \right],
\end{equation*}
and the one for the transverse electromagnetic waves is given by a formula similar to \eqref{6}:
\begin{equation*}
\omega _\textrm{t}^{2}\left( \mathbf{k} \right)=\omega _{0}^{2}+{c^2}{{k}^{2}},
\end{equation*}
where $\omega _0$ is Langmuir oscillation frequency, $r_\textrm{D}$ is Debye radius, $c$  is the speed of light in vacuum.
In  \eqref{3} when replacing sums over the momentum by the integrals and introducing the system volume $V$, atomic density $n\equiv N/V$ and photon density $n_\textrm{ph}\equiv N_\textrm{ph} / V$ and when taking into account \eqref{4} we get the following system of equations:
\begin{equation}%%%%%%%%%%%%%%%%%%% ->chandges an order
\begin{split}\begin{gathered}
  \label{9}
n=\frac{g_{\alpha _1}}{2 \pi ^2 {\hbar }^3}\int\limits_{0}^{\infty }{\rd p}\frac{p^2}{\exp \left[ \frac{{\varepsilon _{\alpha _1}}-{\mu _1}+\left( p^2 /{2m} \right)}{T} \right]\pm 1}
+\frac{g_{\alpha _2}}{2\pi ^2  {\hbar }^3}\int\limits_{0}^{\infty }{\rd p}\frac{p^2}{\exp \left[ \frac{{\varepsilon _{\alpha _2}}-{\mu _2}+\left(  p^2 /{2m} \right)}{T} \right]\pm 1}\,,
\\{{n}_\textrm{ph}}=\frac{g_\textrm{ph}}{{2 \pi ^2}{\hbar ^3}}\int\limits_{0}^{\infty }{\rd p\frac{{p^2}}{\exp \left[ \frac{\hbar {\omega _0}-{\mu ^{*}}+\left( {p^2}/{2m^*} \right)}{T} \right]-1}
} + \frac{g_{\alpha _2}}{2{{\pi }^{2}}{{\hbar }^{3}}}\int\limits_{0}^{\infty }{\rd p}\frac{p^2}{\exp \left[ \frac{\varepsilon_{\alpha _2}-{\mu _2}+\left( {p^2}/{2m} \right)}{T} \right]\pm 1}\,,
\\
{\mu _1}+{\mu ^*}={\mu _2}\,.
\end{gathered}
\end{split}\end{equation}
Here, it was assumed that the photon dispersion relation is given by \eqref{5}. The system of equations \eqref{9} can be used as a starting point when studying the thermodynamic equilibrium of photons with an ideal gas of two-level atoms in a wide temperature range. The system gets rather simplified in some specific cases and conditions of Bose-Einstein condensation can be obtained analytically. For instance, one of such possible situations was considered in \cite{20}, where the conditions of Bose-Einstein condensation of free photons theoretically were received at a high temperature. The term ``high temperature'' means that atomic gas is nondegenerate, thus bosons and fermions behave similarly and there is no need to distinguish the difference between them. As it was mentioned above, this case was specified in detail in \cite{20}; for this reason, in the next section we shall use equations \eqref{9} with the temperature close to the degeneracy temperature of atomic gas components to study the conditions of photon Bose-Einstein condensation {(some other case of photon condensation see in \cite{Sobyanin,Ours})}. In this temperature range, the difference between Fermi-Dirac and Bose-Einstein statistics is very significant. Specifically, bosons can form a BEC below a certain temperature. In the next section we shall find out at what conditions two Bose condensates, i.e., atomic and photonic, can coexist in the system.

To summarize this section, we shall define the densities of atomic components $n_{\alpha _i}\left( T \right) $ where $i=1,2$, the density of free photons $n_\textrm{ph}\left( T \right)$  in the system as follows:
\begin{equation}\begin{split}\begin{gathered}
 \label{10}
 n_{\alpha _i}\left( T \right)\equiv \int{d\mathbf{p \ }}{n_{\alpha _i}}\left( \mathbf{p} \right),
 \qquad i=1,2 ,
\\
{n_\textrm{ph}}\left( T \right)\equiv \int{d\mathbf{p \ }}{n_\textrm{ph}}\left( \mathbf{p} \right), \qquad \phantom{i=1,2,}
\end{gathered}\end{split}\end{equation}
and the atomic distribution function $n_{\alpha _i}\left( \mathbf{p} \right)$ and photon distribution function ${n_\textrm{ph}}\left( \mathbf{p} \right)$:
\begin{equation}\begin{split}\begin{gathered}
 \label{11}
\ {n_{\alpha _i}}\left( \mathbf{p} \right)\equiv \frac{g_{\alpha _i}}{{\left( 2\pi \hbar  \right)}^3}\frac{1}{\exp \left[ \frac{{\varepsilon _{\alpha _i}}-{\mu _i}+\left( {\mathbf{p}^2}/{2m} \right)}{T} \right]\pm 1}\,,
\qquad i=1,2 ,
\\
{n_\textrm{ph}}\left( \mathbf{p} \right)\equiv \frac{g_\textrm{ph}}{\left( 2\pi \hbar  \right)^3} \frac{1}{\exp \left[ \frac{\hbar {\omega _0}-{\mu ^{*}} +\left( {\mathbf{p}^2}/{2m^*} \right)}{T} \right]- 1} \,. \phantom{\qquad i=1,2 ,}
\end{gathered}\end{split}\end{equation}

\section{The coexistence conditions of Bose Einstein condensates of nonexcited atoms and photons }

Three essentially different cases are possible when Bose-Einstein condensation of photons appears in a system at low temperatures. Each case is determined by the type of atoms present in the condensate. In the first case, ground-state atoms form Bose-Einstein condensate; here, for simplicity we shall consider the gas of excited atoms to be nondegenerate. In the second case, excited atoms form Bose-Einstein condensate and (again for simplicity) the gas of ground-state atoms is nondegenerate. In the  third case, all atomic components (gases of excited and non-excited atoms) form  Bose-Einstein condensate.

When studying the listed cases let us stick to the general system of equations rearranged by taking into account the fact that an atomic subsystem consists of two-level bose-atoms:
\begin{equation} \begin{split}\begin{gathered}
 \label{12}
n=\frac{{{g}_{{{\alpha }_{1}}}}}{2{{\pi }^{2}}{{\hbar }^{3}}}\int\limits_{0}^{\infty }{\rd p}\frac{{{p}^{2}}}{\exp \left[ \frac{{{\varepsilon }_{{{\alpha }_{1}}}}-{{\mu }_{1}}+\left( {{{p}^{2}}}/{2m}\right)}{T} \right]- 1}
+
\frac{{{g}_{{{\alpha }_{2}}}}}{2{{\pi }^{2}}{{\hbar }^{3}}}\int\limits_{0}^{\infty }{\rd p}\frac{{{p}^{2}}}{\exp \left[ \frac{{{\varepsilon }_{{{\alpha }_{2}}}}-{{\mu }_{2}}+\left( {{{p}^{2}}}/{2m} \right)}{T} \right]- 1}\,,
\\{{n}_\textrm{ph}}=\frac{{{g}_\textrm{ph}}}{2{{\pi }^{2}}{{\hbar }^{3}}}\int\limits_{0}^{\infty }{\rd p\frac{{{p}^{2}}}{\exp \left[ \frac{\hbar {{\omega }_{0}}-{{\mu }^{*}}+\left( {{{p}^{2}}}/{2m} \right)}{T} \right]-1}}
+
\frac{{{g}_{{{\alpha }_{2}}}}}{2{{\pi }^{2}}{{\hbar }^{3}}}\int\limits_{0}^{\infty }{\rd p}\frac{{{p}^{2}}}{\exp \left[ \frac{{{\varepsilon }_{{{\alpha }_{2}}}}-{{\mu }_{2}}+\left( {{{p}^{2}}}/{2m} \right)}{T} \right]- 1}\,,
\\[1pt]
{{\mu }_{1}}+{{\mu }^{*}}={{\mu }_{2}}\,.
\end{gathered}\end{split}\end{equation}
Let us remind that \eqref{9} assumes that the system is in thermodynamic equilibrium. Note that according to \eqref{10} and \eqref{11}, two first equations \eqref{12} can be written as follows:
\begin{equation}\begin{split}\begin{gathered}
\label{13}
n=\int{\rd\mathbf{p \ }}{{n}_{{{\alpha }_{1}}}}\left( \mathbf{p} \right)+\int{\rd\mathbf{p}}{{n}_{{{\alpha }_{2}}}}\left( \mathbf{p} \right),
\\
{{n}_\textrm{ph}}=\int{\rd\mathbf{p \ }}{{n}_\textrm{ph}}\left( \mathbf{p} \right)+\int{\rd\mathbf{p}}{{n}_{{{\alpha }_{2}}}}\left( \mathbf{p} \right).
\end{gathered}\end{split}\end{equation}
First we study the case of Bose-Einstein condensates formed by photons and ground-state atoms, whereas the gas of excited atoms is non-degenerate. Since the gas is considered to be non-degenerate, the chemical potential of excited atoms $\mu _2$ satisfies the condition:
\begin{equation}
\label{14}
\exp \left( \frac{{{\varepsilon }_{{{\alpha }_{2}}}}-{{\mu }_{2}}}{T} \right)\gg 1\ .
\end{equation}
The previous inequality makes it easy to calculate the integrals in the last expressions of the first and second equations of a system \eqref{12} and to get the next equations set:
\begin{equation}\begin{split}\begin{gathered}
 \label{15}
n=\frac{{{g}_{{{\alpha }_{1}}}}}{2{{\pi }^{2}}{{\hbar }^{3}}}\int\limits_{0}^{\infty }{\rd p}\frac{{{p}^{2}}}{\exp \left[ \frac{{{\varepsilon }_{{{\alpha }_{1}}}}-{{\mu }_{1}}
+\left( {{{\mathbf{p}}^{2}}}/{2m} \right)}{T} \right]-1}
+
{{g}_{{{\alpha }_{2}}}}{{\left( \frac{mT}{2\pi {{\hbar }^{2}}} \right)}^{3/2}} \exp \left[ {\left( {{\mu }_{2}}-{{\varepsilon }_{{{\alpha }_{2}}}} \right)}/{T} \right]\,,
\\
{{n}_\textrm{ph}}=\frac{{{g}_\textrm{ph}}}{2{{\pi }^{2}}{{\hbar }^{3}}}\int\limits_{0}^{\infty }{\rd p\frac{{{p}^{2}}}{\exp \left[ \frac{\hbar {{\omega }_{0}}-{{\mu }^{*}}+\left( {{{p}^{2}}}/{2{{m}^{*}}}\right)}{T} \right]-1}}
+{{g}_{{{\alpha }_{2}}}}{{\left( \frac{mT}{2\pi {{\hbar }^{2}}} \right)}^{3/2}} \exp \left[ {\left( {{\mu }_{2}}-{{\varepsilon }_{{{\alpha }_{2}}}} \right)}/{T} \right]\,,
\\[1pt  ]
{{\mu }_{1}}+{{\mu }^{*}}={{\mu }_{2}}\,.
\end{gathered}\end{split} \end{equation}
Note that all components of the system studied are ideal gases. Consequently, a photonic component and a component of atoms in the ground state are required to satisfy the equalities (see in this regard, for example, \cite{24}) for Bose condensate to appear in the system:
\begin{equation}
\label{16}
{{\mu }^{*}}\left| _{T\leqslant T_\textrm{c}^{*}} \right.=\hbar {{\omega }_{0}}\,, \qquad
{{\mu }_{1}}\left| _{T\leqslant {{T}_\textrm{c}}} \right.={{\varepsilon }_{1}}\,.
\end{equation}
There were introduced parameters $T_\textrm{c}$, i.e., the condensation temperature of the ground state atomic gas and $T_\textrm{c}^{*}$, i.e.,  the condensation temperature of the photon gas, correspondingly. As a result of expression \eqref{16} and the last equation in \eqref{15} transformation, we get chemical potential $\mu _2$ formula:
\begin{equation}
\label{17}
{{\mu }_{2}}={{\varepsilon }_{1}}+\hbar {{\omega }_{0}}\,.
\end{equation}
Taking into account \eqref{16}, \eqref{17}, \eqref{13}, the first two equations of \eqref{15} can also be rearranged as:
\begin{equation}\begin{split}\begin{gathered}
\label{18}
n=\int{\rd\mathbf{p}}{{n}_{{{\alpha }_{1}}}}\left( \mathbf{p} \right)+{{g}_{{{\alpha }_{2}}}}{{\left( \frac{mT}{2\pi {{\hbar }^{2}}} \right)}^{3/2}}\exp \left( -{\Delta }/{T}\right) ,
\\
{{n}_\textrm{ph}}=\int{\rd\mathbf{p}}{{n}_\textrm{ph}}\left( \mathbf{p} \right)+{{g}_{{{\alpha }_{2}}}}{{\left( \frac{mT}{2\pi {{\hbar }^{2}}} \right)}^{3/2}}\exp \left( -{\Delta }/{T} \right),
\end{gathered}\end{split} \end{equation}
where $n_{\alpha _1}\left( \mathbf{p} \right)$ , ${{n}_\textrm{ph}}\left( \mathbf{p} \right)$ were defined in \eqref{10}, \eqref{11} and symbol $\Delta$ means:
\begin{equation}
\label{19}
\Delta \equiv {{\varepsilon }_{{{\alpha }_{2}}}}-{{\varepsilon }_{{{\alpha }_{1}}}}-\hbar {{\omega }_{0}}={{\varepsilon }_{{{\alpha }_{2}}}}-{{\mu }_{2}}\,.
\end{equation}
In the case being studied, the value of $\Delta $ should be greater than zero; the inequality $\Delta >0$ is required because excited atoms are non-degenerated, see \eqref{14}.

Equations \eqref{18}, \eqref{19} are the ones to be studied as the initial equations to define the characteristics of conditions under which photons and non-excited atoms BEC coexist. The density of photons ${{n}_\textrm{ph}}\left( \mathbf{p} \right)$ and atoms ${{n}_{{{\alpha }_{1}}}}\left( \mathbf{p} \right)$ distribution function below the transition temperatures ${{T}_\textrm{c}}$ and $T_\textrm{c}^{*}$ over the momentum can be represented as follows (in this regard see, for example \cite{24}):
\begin{equation}\begin{split}\begin{gathered}
\label{20}
{{n}_{{{\alpha }_{1}}}}\left( \mathbf{p} \right)=\frac{{{g}_{{{\alpha }_{1}}}}}{{{\left( 2\pi \hbar \right)}^{3}}}{{\left[ \exp \left( \frac{{{\mathbf{p}}^{2}}}{2mT} \right)-1 \right]}^{-1}}
+n_{{{\alpha }_{1}}}^{0}\left( T \right)\delta \left( \mathbf{p} \right), \qquad
 T\leqslant {{T}_\textrm{c}}\,,
\\
{{n}_\textrm{ph}}\left( \mathbf{p} \right)=\frac{{{g}_\textrm{ph}}}{{{\left( 2\pi \hbar \right)}^{3}}}{{\left[ \exp \left( \frac{{{\mathbf{p}}^{2}}}{2{{m}^{*}}T} \right)-1 \right]}^{-1}}
+n_\textrm{ph}^{0}\left( T \right)\delta \left( \mathbf{p} \right), \qquad  T\leqslant T_\textrm{c}^{*},
\end{gathered}\end{split} \end{equation}
where $n_{{{\alpha }_{1}}}^{0}\left( T \right)$ is BEC density of atoms, $n_\textrm{ph}^{0}\left( T \right)$ is BEC density of free photons. Let us put \eqref{20} into \eqref{18} to obtain expressions for such densities:
\begin{equation}\begin{split}\begin{gathered}
\label{21}
n_{\alpha _1}^{0}\left( T \right)=n-{{\left( \frac{\!mT}{2\pi {{\hbar }^{2}}} \right)}^{{3}/{2}}} {{g}_{{{\alpha }_{1}}}}\zeta \left( {3}/{2} \right)
+{{\left( \frac{\!mT}{2\pi {{\hbar }^{2}}} \right)}^{{3}/{2}}}{{g}_{{{\alpha }_{2}}}}\exp\left(-{\Delta }/{T}\right) ,
\\
n_\textrm{ph}^{0}\left( T \right)={{n}_\textrm{ph}}-{{g}_\textrm{ph}}{{\left( \frac{{{m}^{*}}T}{2\pi {{\hbar }^{2}}} \right)}^{\!{3}/{2}}}\!\zeta\left( {3}/{2}\right)
-{{g}_{{{\alpha }_{2}}}}{{\left( \frac{mT}{2\pi {{\hbar }^{2}}} \right)}^{{3}/{2}}}\exp\left( -{\Delta }/{T}\right),
\end{gathered}\end{split} \end{equation}
where $\zeta \left( x \right)$ is Riemann zeta function. We emphasize that BEC disappear in the transition point; this fact infers the following:
\begin{equation}
\label{22}
n_{{{\alpha }_{1}}}^{0}\left( {{T}_\textrm{c}} \right)=0, \qquad n_\textrm{ph}^{0}\left( T_\textrm{c}^{*} \right)=0.
\end{equation}
Thus, \eqref{22} together with \eqref{21} should be used to define transition temperatures $T_\textrm{c}$ and $T_\textrm{c}^{*}$. To analyze the cases possible for the first equation in \eqref{21} let us regard the temperature to be equal to ${{T}_\textrm{c}}$, i.e., ground-state atoms condensation temperature; for the second equation in \eqref{21} let us regard the temperature to be equal to $T_\textrm{c}^{*}$, i.e., photons condensation temperature. As a result, we get:
\begin{equation}\begin{split}\begin{gathered}
\label{23}
n={{\left( \frac{m{{T}_\textrm{c}}}{2\pi {{\hbar }^{2}}} \right)}^{{3}/{2}\!}}\big[ {{g}_{{{\alpha }_{1}}}}\zeta \left( {3}/{2}\!\right)+
{{g}_{{{\alpha }_{2}}}}\exp \left( -{\Delta }/{{{T}_\textrm{c}}}\right) \big],
\\
{{n}_\textrm{ph}}={{g}_\textrm{ph}}{{\left( \frac{{{m}^{*}}T_\textrm{c}^{*}}{2\pi {{\hbar }^{2}}} \right)}^{{3}/{2}}}\!\!\zeta \left( {3}/{2}\right)
+{{g}_{{{\alpha }_{2}}}}{{\left( \frac{mT_\textrm{c}^{*}}{2\pi {{\hbar }^{2}}} \right)}^{3/2}}\exp \left( -{\Delta }/{T_\textrm{c}^{*}}\!\right).
\end{gathered}\end{split} \end{equation}
It is easy to see that equations \eqref{23} are transcendental; they do not have analytical solution~--- only the numerical one. Nevertheless, in some cases analytical solution can be obtained.

Let us analyze the first equation in \eqref{23} when temperatures are supposed to be low: it means that inequality $\exp \left( {-\Delta }/{T_\textrm{c}}\right)\ll 1$ [or $\left( {T_\textrm{c}}/{\Delta } \right)\ll 1$] is valid. When temperatures are low the first equation in \eqref{23} can be solved by means of perturbation theory using parameter $\exp \left( {-\Delta }/{T_\textrm{c}}\right)\ll 1$ and we get:
\begin{equation}\begin{split}\begin{gathered}
\label{24}
{{T}_\textrm{c}}\approx \frac{2\pi {{\hbar }^{2}}}{m}{{\left[ \frac{n}{\xi \left( {3}/{2} \right){{g}_{{{\alpha }_{1}}}}} \right]}^{{2}/{3}}}\left[ 1-
\frac{2}{3}\frac{{{g}_{{{\alpha }_{2}}}}}{{{g}_{{{\alpha }_{1}}}}}\frac{1}{\xi \left( {3}/{2}\right)}\exp \left( -\frac{\Delta }{{{T}_\textrm{c}}} \right) \right].
\end{gathered}\end{split} \end{equation}
The first order of perturbation theory gives us the following equation using the method of successive approximation:
\begin{equation}%\begin{split}\begin{gathered}
 \label{25}
{{T}_\textrm{c}}\approx \!\frac{2\pi {{\hbar }^{2}}}{m}{{\left[ \frac{n}{\zeta \left( 3/2\right){{g}_{{{\alpha }_{1}}}}} \right]}^{2/3}}
\left( 1-
\frac{2}{3}\frac{{{g}_{{{\alpha }_{2}}}}}{{{g}_{{{\alpha }_{1}}}}}\frac{1}{\zeta \left( 3/2 \right)}
\exp \left\{-\frac{m\Delta \left[\zeta \left( {3}/{2}\right){g_{\alpha _1}}\right]^{3/2}}{2\pi {\hbar }^2 n^{3/2}}{ {} }\right\} \right).
%\end{gathered}\end{split}
\end{equation}
This equation gives adaptability criteria for the low temperature approximation:
\begin{equation}
\label{26}
\frac{m\Delta }{2\pi {{\hbar }^{2}}}{{\left( \frac{n}{\zeta \left( {3}/{2}\right){{g}_{{{\alpha }_{1}}}}} \right)}^{{-3}/{2}}}\!\gg 1.
\end{equation}
We have a quite different case when defining the condensation temperature of free photons. Let us write the second equation of the system \eqref{23} as follows:
\begin{equation}\begin{split}\begin{gathered}
\label{27}
{{n}_\textrm{ph}}={{\left( \frac{mT_\textrm{c}^{*}}{2\pi {{\hbar }^{2}}} \right)}^{3/2}}\left[ {{g}_\textrm{ph}}\zeta \left( {3}/{2} \right){{\left( \frac{{{m}^{*}}}{m} \right)}^{{3}/{2}}}+
{{g}_{{{\alpha }_{2}}}}\exp \left( -{\Delta }/{T_\textrm{c}^{*}}\right) \right].
\end{gathered}\end{split} \end{equation}
It should be pointed out that the ${m^{*}}/m $ value, i.e., the relation of photon mass to the mass of atom, is very small. (A very different situation was treated in detail in \cite{another one} where full solution for the case of two atomic species forming a molecule was given). For instance, in \cite{18} the authors estimated photon mass as $6.7\times {10}^{-33}$ g (see also \cite{20}). Even for lithium, that relation equals the order-of-magnitude $ \left( {m^{*}}/m \right)\sim 10^{-10}$. This circumstance gives us two possible cases determined by two inequalities:
\begin{equation}\begin{split}\begin{gathered}
\label{28}
{{\left( \frac{{{m}^{*}}}{m} \right)}^{{3}/{2}}}\!\!\gg \exp \left( {-\Delta }/{T_\textrm{c}^{*}}\right),
\\
{{\left( \frac{{{m}^{*}}}{m} \right)}^{{3}/{2}}}\!\!\ll \exp \left( {-\Delta }/{T_\textrm{c}^{*}} \right).
\end{gathered}
\end{split}\end{equation}
If there are lithium atoms in the system, the values ${\left( {m^*}/m \right)}^{3/2}$ and $\exp \left( {-\Delta }/{T_\textrm{c}^{*}}\right)$ will be of the same order of magnitude at $\Delta \sim 30 \ T_\textrm{c}^{*}$
\begin{equation}
\label{29}
{{\left( \frac{{{m}^{*}}}{m} \right)}^{{3}/{2}}}\sim \exp \left( {-\Delta }/{T_\textrm{c}^{*}} \right), \qquad \Delta \sim 30 \ T_\textrm{c}^{*}.
 \end{equation}
To analyze the situation to which the validity of the first or the second equality in \eqref{28} may lead, let us perform as follows: we divide the photon density (when system temperature is $T_\textrm{c}^{*}$) by atoms density (when system temperature is $T_\textrm{c}$). As a result, we have:
\begin{equation}
\label{30}
\frac{{{n}_\textrm{ph}}}{n}=\!\!{{\left( \frac{T_\textrm{c}^{*}}{{{T}_\textrm{c}}} \right)}} \frac{
 {{{g}_\textrm{ph}}\zeta \left( {3}/{2}\right){{\left( {m^*}/{m} \right)\!}^{{3}/{2}\!}}\!+\!{{g}_{{{\alpha }_{2}}}}\exp \left( {-\Delta }/{T_\textrm{c}^{*}} \right)}}{{{{g}_{{{\alpha }_{1}}}}\zeta \left( {3}/{2}\right)+{{g}_{{{\alpha }_{2}}}}\exp \left( {-\Delta }/{{{T}_\textrm{c}}} \right)}}\,.
\end{equation}
All important characteristics of the system, i.e., condensation temperatures $T_\textrm{c}^{*}$, $T_\textrm{c}$ and the ratio of the masses $m^*/m$ and of the densities ${n_\textrm{ph}}/{n}$, are included in \eqref{30}. As it was mentioned above, condensation temperature of photons was assumed to be far below degeneration temperature of atomic gas in \cite{20}. Here, quite different case is studied: atomic and photonic condensates were considered as coexisting ones. For definiteness, we have attributed the condensation temperature of photons to $T_\textrm{c}^{*}$ and assigned it to be below atomic condensation temperature $T_\textrm{c}^{*} \lesssim T_\textrm{c}$. If the first inequality in \eqref{28} is valid, we can neglect its compounds with exponents in \eqref{30}, and in the main approximation we shall get:
\begin{equation}{
\label{31}
\frac{{{n}_\textrm{ph}}}{n}\approx \frac{{{g}_\textrm{ph}}}{{{g}_{{{\alpha }_{1}}}}}\,{{\left( \frac{{{m}^{*}}}{m} \right)}^{{3}/{2}}}.
} \end{equation}
When atomic gas is helium, the latter ratio is estimated to be $(n_\textrm{ph}/n)\sim 10^{-15}$. In other words, the fulfillment of the first case in \eqref{28} implied that the amount of photons is negligibly small in comparison with the number of atoms in the system. This particular value, i.e., the density of photons, defines the maximum possible photon condensate density in the system (according to \eqref{21}). This circumstance makes the case (defined by first formula in \eqref{28}) not worth studying in this paper, because this is not within the scope of priorities, i.e., to obtain photonic BEC, i.e., which was stated at the beginning of our research.

We shall definitely get a more interesting case if the second condition in \eqref{28} is valid where it becomes more promising to obtain BEC. From \eqref{12} one can get:
\begin{equation}\begin{split}\begin{gathered}
\label{32}
\frac{{{n}_\textrm{ph}}}{n}\approx \frac{{{g}_{{{\alpha }_{2}}}}\exp \left( -{\Delta }/{T_\textrm{c}^{*}} \right)}{{{g}_{{{\alpha }_{1}}}}\zeta \left( {3}/{2}\right)}\gg \frac{{{g}_\textrm{ph}}}{{{g}_{{{\alpha }_{1}}}}}\,{{\left( \frac{{{m}^{*}}}{m} \right)}^{{3}/{2}}},
\qquad
\exp \left( -\frac{\Delta }{T_\textrm{c}^{*}} \right)\ll 1.
\end{gathered}
\end{split}\end{equation}
According to the second condition in \eqref{32}, the inequality ${n_\textrm{ph}}/{n}\ll $ is maintained. Despite that fact, the ratio \eqref{32} gives us the ability to observe BEC coexistence of atoms and photons at ultralow temperatures at higher values of photon density than formula \eqref{31} provides and taking into account \eqref{29}, \eqref{27}, \eqref{32}, we can find the following equation for the critical temperature:
\begin{equation}\begin{split}\begin{gathered}
\label{33}
{{n}_\textrm{ph}}={{g}_{{{\alpha }_{2}}}}{{\left( \frac{mT_\textrm{c}^{*}}{2\pi {{\hbar }^{2}}} \right)}^{3/2}}\exp \left( -{\Delta }/{T_\textrm{c}^{*}} \right),
\qquad
{1}/{30}< {T_\textrm{c}^{*}}/{\left| \Delta \right|}\ll1,
\end{gathered}
\end{split}\end{equation}
or write it as follows:
\begin{equation}\begin{split}\begin{gathered}
\label{34}
T_\textrm{c}^{*}=\frac{2\pi {{\hbar }^{2}}}{m}{{\left( \frac{{{n}_\textrm{ph}}}{{{g}_{{{\alpha }_{2}}}}} \right)}^{{2}/{3}\;}}\exp \left( \frac{2}{3}\frac{\Delta }{T_\textrm{c}^{*}} \right)
={{T}_\textrm{c}}{{\left[ \frac{{{g}_{{{\alpha }_{1}}}}{{n}_\textrm{ph}}}{\zeta \left( {3}/{2}\right){{g}_{{{\alpha }_{2}}}}n} \right]}^{{2}/{3}\;}}\exp \left( \frac{2}{3}\frac{\Delta }{T_\textrm{c}^{*}} \right),
 \\[1pt]
{1}/{30}< {T_\textrm{c}^{*}}/{\left| \Delta \right|}\ll1.
\end{gathered}
\end{split}\end{equation}
It is easy to see in this case that the critical temperature dependence on the density $n_\textrm{ph}$ is not power-behaved and essentially non-typical for transition points with BEC in ideal gases (see \cite{24} for example); this behavior was first noticed in \cite{20}. The condition $T_\textrm{c}^* / T_\textrm{c} \lesssim 1$ taking into consideration \eqref{25}, \eqref{33} can be represented as follows:
\begin{equation}{
\label{35}
\frac{T_\textrm{c}^{*}}{{{T}_\textrm{c}}}={{\left[ \frac{{{g}_{{{\alpha }_{1}}}}{{n}_\textrm{ph}}}{\zeta \left( {3}/{2}\right)\!{{g}_{{{\alpha }_{2}}}}n} \right]}^{{2}/{3}}}\!\exp \left( \frac{2}{3}\frac{\Delta }{T_\textrm{c}^{*}} \right)\lesssim1
} \end{equation}
and gives us the condition under which the situation being studied is possible:
\begin{equation} {
\label{36}
{{\left[ \frac{{{n}_\textrm{ph}}}{n}\exp \left( \frac{\Delta }{T_\textrm{c}^{*}} \right) \right]}^{{2}/{3}}}\lesssim1
} \end{equation}
and this does not contradict with \eqref{32}.

Equations \eqref{33} [or \eqref{34}] are both transcendental and do not have any analytical solution; it is easy to verify that a solution exists only when the ratio of densities satisfies ${n_\textrm{ph}}/n \ll 1$ [condition \eqref{32} also fulfils], because the temperature belongs to the next $1/30 < (T_\textrm{c} ^* / \Delta) \ll 1$ temperature range. For example, when one supposes $(T_\textrm{c} ^*/\Delta) \approx 0.1$, $\exp (-{\Delta}/{T_\textrm{c}^*}) \approx 4.5 \times 10^{-5}$ and consequently takes into consideration \eqref{33}, \eqref{34} and if the fact that $(T_\textrm{c} / T_\textrm{c}^* \sim 1)$ is finally true, we get the following:
\begin{equation}
\label{37}
\frac{{{n}_\textrm{ph}}}{n}\sim \exp \left( -\frac{\Delta }{T_\textrm{c}^{*}} \right)\approx 4.5\times{{10}^{-5}}.
\end{equation}
In other words, if particle density satisfies $n \sim 10^{12}$--$10^{14}$~cm$^{-3}$ (these densities are typical of the experiments performed at ultralow temperatures, see \cite{1,2,3}), our studied case may be expected to occur when total density of photons satisfies $ n_\textrm{ph} \sim 10^{7}$--$ 10^{9}$~cm$^{-3} $.

\begin{figure}[!t]
\centerline{\includegraphics[width=0.45\textwidth]{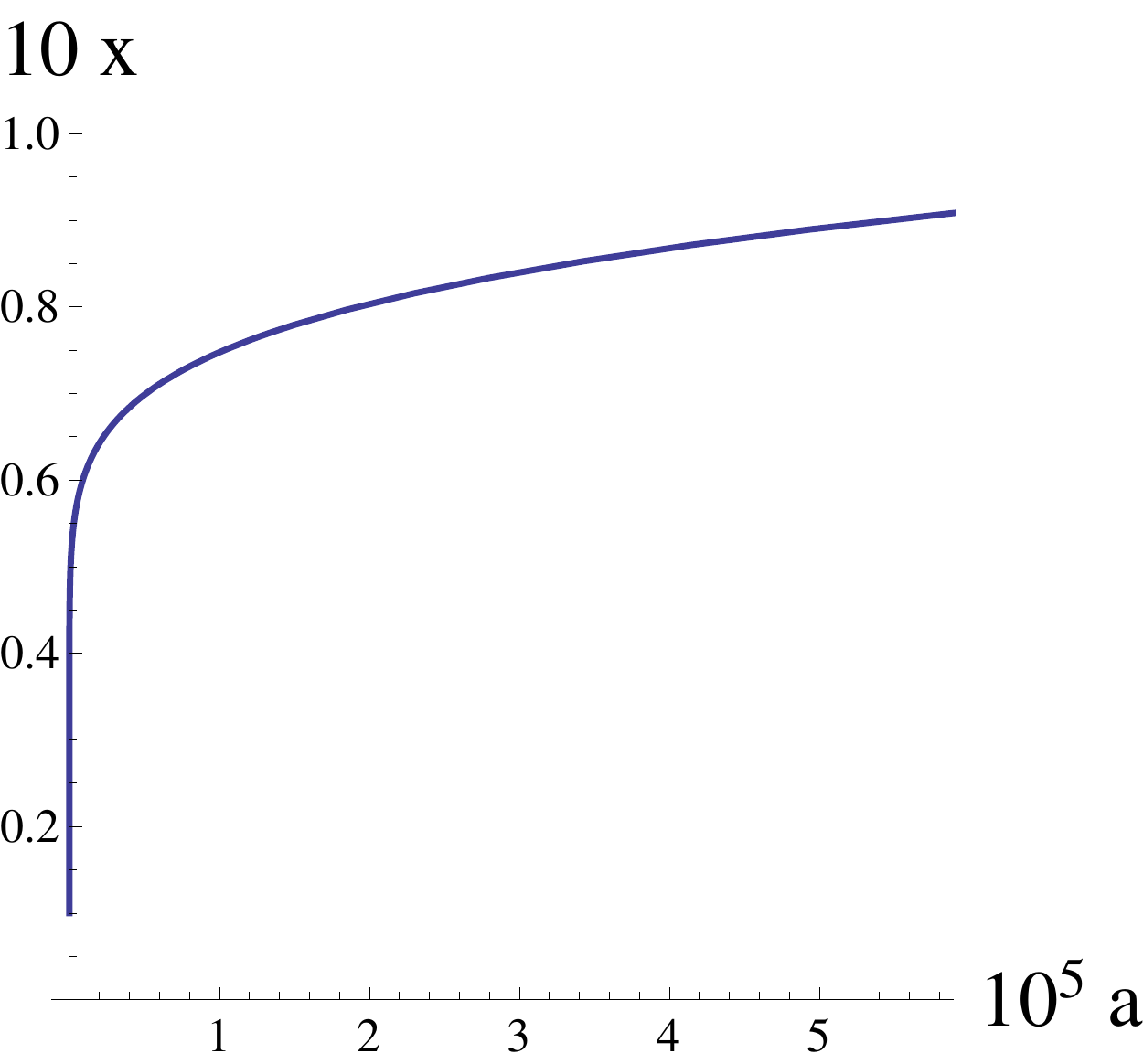}}
\caption{
The transition temperature dependence on $\Delta$   and overall photonic density.}
\label{first}
 \end{figure}

Such photon transition temperature under which BEC appears may be estimated from figure~\ref{first} where the numerical solution of \eqref{34} has been shown. The dimensionless quantities have been introduced as follows:
\begin{equation}
 \label{38}\
a\equiv \frac{1}{\Delta }\frac{2\pi {{\hbar }^{2}}}{m}{{\left( \frac{{{n}_\textrm{ph}}}{{{g}_{{{\alpha }_{2}}}}} \right)}^{{2}/{3} }} ,\qquad x\equiv \frac{T_\textrm{c}^{*}}{\Delta },
\end{equation}
and equation \eqref{38} has been transformed as follows:
\begin{equation}
 \label{39}
 x=a\exp \left( \frac{2}{3x} \right) .	
\end{equation}

Approximated formulae for densities of photonic $n_\textrm{ph}^0 (T)$ and atomic $n_{\alpha_1}^0 (T)$ components in the system being studied can be obtained from \eqref{21} using \eqref{25}, \eqref{28}, \eqref{31}--\eqref{33}; for the first order of smallness by $\exp (-\Delta/T)$, one can obtain:
\begin{equation}\begin{split}\begin{gathered}
\label{40}
n_{{{\alpha }_{1}}}^{0}\left( T \right)\approx n\Bigg\{ 1-{{\left( \frac{T}{{{T}_\textrm{c}}} \right)}^{{3}/{2} }}
-\frac{{{g}_{{{\alpha }_{2}}}}}{\zeta \left( {3}/{2} \right){{g}_{{{\alpha }_{1}}}}}{{\left( \frac{T}{{{T}_\textrm{c}}} \right)}^{{3}/{2} }}
\Bigg[ \exp \left( -\frac{\Delta }{T} \right)-\exp \left( -\frac{\Delta }{{{T}_\textrm{c}}} \right) \Bigg] \Bigg\},
\\
n_\textrm{ph}^{0}\left( T \right)\approx {{n}_\textrm{ph}}\left\{ 1-{{\left( \frac{T}{T_\textrm{c}^{*}} \right)}^{{3}/{2} }}\!\!\exp \left[ -\Delta \left( \frac{1}{T}-\frac{1}{T_\textrm{c}^{*}} \right) \right] \right\}.
\\
\end{gathered} \end{split}\end{equation}

According to  \eqref{40} photon BEC occurs rapidly  together with temperature
  decrease (we call it  an abrupt  condensation mode).  This can be easily seen when calculating the derivative $\partial{n_\textrm{ph}^0} / {\partial T}$ value at the following temperatures $T \lesssim T_\textrm{c}^* $:
\begin{equation}{
\label{41}
{{\left. \frac{\partial n_\textrm{ph}^{0}\left( T \right)}{\partial T} \right|}_{T\lesssim T_\textrm{c}^{*}}}\approx-\frac{{{n}_\textrm{ph}}}{T_\textrm{c}^{*}}
\frac{\Delta}{T_\textrm{c}^{*}}\,, \qquad \frac{\Delta }{T_\textrm{c}^{*}}\gg 1.
} \end{equation}
It should be emphasized that we do not mean that by lowering the temperature  the thermalization in the system and photonic  BEC appearance could happen promptly; here, we mean that the slightest decrease of  the temperature can cause a significant  increase of the photon BEC density.

\begin{figure}
\centerline{\includegraphics[width=0.55\textwidth] {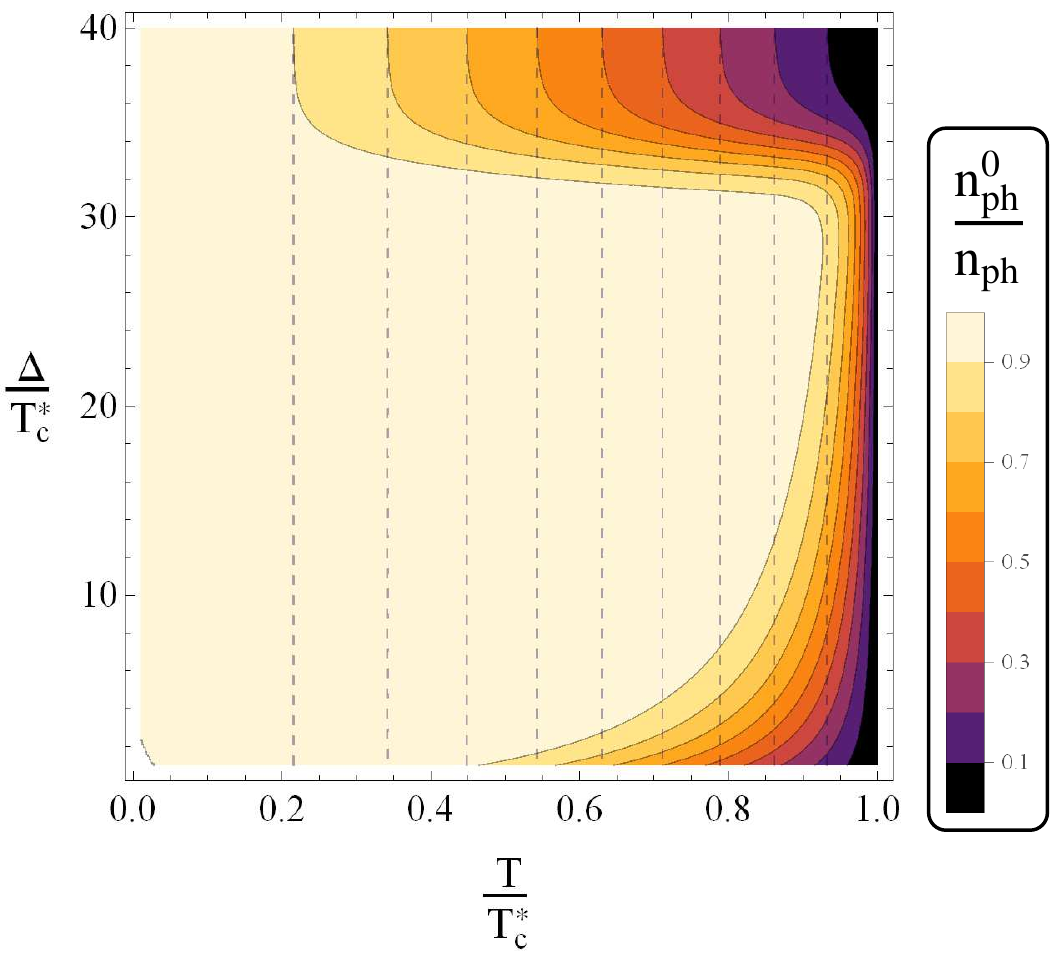} }
\caption{(Color online) The demonstration of the possibility of an abrupt condensation of photons being in equilibrium with ultracold atomic gas.
}
\label{second}
\end{figure}

The number of excited atoms decreased in the same manner: one may obtain it from formulae \eqref{18}, \eqref{19} and when taking into account the equation \eqref{13} for such  density of atoms $n_{\alpha_2} (T)$, we get the following:
\begin{equation}{
\label{42}
{{n}_{{{\alpha }_{2}}}}\left( T \right)={{g}_{{{\alpha }_{2}}}}{{\left( \frac{mT}{2\pi {{\hbar }^{2}}} \right)}^{\!\!3/2}}\!\!\exp \left( -{\Delta }/{T} \right).\
} \end{equation}
The opposite statement is also correct: when temperature of the system is decreasing the state with BEC disappears and photons ``captured'' by atoms are being emitted. The statement about an abrupt  character of condensation is fully illustrated in figure~\ref{second}: it expresses such a normalized density $n_\textrm{ph}^0 / n_\textrm{ph}$ behavior that depends on dimensionless parameters: $T/T_\textrm{c}^*$ and $\Delta/T_\textrm{c}^*$ that is responsible for low-temperature approximation. There we used a precise expression for a condensate density of photons $n_\textrm{ph}^0 $:
\begin{equation}\begin{split} \begin{gathered}
\label{43}
\frac{n_\textrm{ph}^{0}\left( T \right)}{{n}_\textrm{ph}}= 1-
{{\left( \frac{T}{T_\textrm{c}^{*}} \right)}^{3/2}}
\frac{{{g}_\textrm{ph}}\zeta \left( {3}/{2}\right){{\left( {{{m}^{*}}}/{m} \right)}^{{3}/{2}}}+{{g}_{{{\alpha }_{2}}}}\exp \left( -{\Delta }/{T} \right)}{{{g}_\textrm{ph}}\zeta \left( {3}/{2}\right){{\left( {{{m}^{*}}}/{m} \right)}^{{3}/{2}}}+{{g}_{{{\alpha }_{2}}}}\exp \left( -{\Delta }/{T_\textrm{c}^{*}} \right)} \,,
\end{gathered}\end{split}\end{equation}
that could be easily obtained from the second inequality \eqref{21} when using the second equation in \eqref{23}. On getting the value of $n_\textrm{ph}^0 / n_\textrm{ph}$, we suggest that condensation temperatures are comparable $T \sim T_\textrm{c}^* $ and $T \lesssim T_\textrm{c}^* $.
Figure~\ref{second} was performed in the same way as, for example, rock massif mapping in cartography, but here the higher altitude is indicated with lighter color. Solid line curves in the figure are contour lines that join the dots with the same value of $n_\textrm{ph}^0 / n_\textrm{ph}$ (like isopleths in map making). One may see in this contour graph that in a certain range of $T/T_\textrm{c}^*$ and $\Delta / T_\textrm{c}^*$ values, the abrupt condensation of photons is possible: when nondimensional temperature $T/T_\textrm{c}^* $ decreases slightly, the value of $n_\textrm{ph}^0 / n_\textrm{ph}$ almost immediately grows up to its maximum (in mapping terms it looks like a vast ``plateau''). This abrupt character of condensation of photons corresponds to an abrupt decrease of the population of non-excited atoms according to \eqref{42}.

The area of the contour graph with $ n_\textrm{ph}^0 / n_\textrm{ph} > 0.9 $ is the lightest (where normalized density $n_\textrm{ph}^0 / n_\textrm{ph}$ is the highest). For instance, if $\Delta / T_\textrm{c}^* \approx 20$, the latter inequality is valid for $T/T_\textrm{c}^* \lesssim 0.9$, and if $\Delta / T_\textrm{c}^* \approx 10$, the nondimensional temperature needs to be decreased to $ T/T_\textrm{c}^* \lesssim 0.85$ to satisfy that inequality.

One may notice from approximated formulae \eqref{40}, \eqref{41} that the illustrated abrupt character of condensation does not occur for all parameters $\Delta$ that satisfy ${\Delta}/{T_\textrm{c}^*} \gg1$: from ${\Delta}/{T_\textrm{c}^*} \approx 30$ to higher ${\Delta}/{T_\textrm{c}^*}$ values, the condensation law becomes
``standard'' as for ideal gas~--- when photon BEC density obtains power-law behaved dependency on a temperature
\begin{equation*}
n_\textrm{ph}^{0}\left( T \right)\approx {{n}_\textrm{ph}}\left[ 1-{{\left( \frac{T}{T_\textrm{c}^{*}} \right)}^{{3}/{2}}} \right].
\end{equation*}
The power-law mode of condensation is mapped with dashed lines in figure~\ref{second}; it is evident that in the upper area of the graph (when $\Delta / T_\textrm{c}^* > 30$) the abrupt type of condensation switches to rather different one~--- power-behaved type of condensation.

Obtained expression \eqref{40}, \eqref{43} allows us to reach some other intriguing conclusions. It is easy to see that when temperatures are subjected to the condition:
\begin{equation} {
\label{44}
\frac{T}{{{T}_\textrm{c}}}\ll 1 ,\qquad \frac{T_{}^{*}}{{{T}_\textrm{c}}}<1 , \qquad \frac{{{T}_\textrm{c}}}{\Delta }\ll 1
} \end{equation}
almost all photons are in BEC. Besides, almost all atoms remain unexcited with the set of quantum numbers ${{\alpha }_{1}}$ and form BEC. This situation could be interpreted as ``stopped light'' in BEC, but it is different from the known one. Let us remind that appearance of ``stopped light'' phenomenon was first shown in \cite{9,10,11} and let us  clarify why we used such a term. It is known that group velocity ${{v}_\textrm{g}}$ of electromagnetic waves propagation through some matter can be defined as:
\begin{equation*}
{{v}_\textrm{g}}\equiv \frac{\rd\omega \left( k \right)}{\rd k}\,.
\end{equation*}
When using ``relativistic'' dispersion law \eqref{6} for small wave vectors [see \eqref{7}], one can get the following:
\begin{equation}
\label{45}
{{v}_\textrm{g}}=\frac{vk}{\sqrt{\omega _{0}^{2}+{{v}^{2}}{{k}^{2}}}}\approx \frac{vk}{{{\omega }_{0}}}\,,
\end{equation}
where $v$ is phase velocity of light in the matter. The expression similar to \eqref{45} that was also supposed to apear in this paper will be also valid in case of quadratic photons dispersion law. Zero momentum photons form BEC. It means that their wave vector is zero and correspondingly to \eqref{45} their group velocity is zero. This is the situation that was foreseen when we stated a possibility of stopped light phenomena in studied system: according to formulas \eqref{40}--\eqref{45}, photons could be ``captured'' in BEC of atomic gas with their possible following transition into coherent state, because each photon in the condensate has
${{\left. \omega \left( k \right) \right|}_{k=0}}={{\omega }_{0}}$.
We note that this case does not only concern the light itself but also electromagnetic waves in general.

\section{On possibility of BEC coexistence in photon component and excited atoms subsystem}

Let us now study the case when BEC is formed not only by photons but also by excited atoms; for simplicity, ground state atoms are assumed to be nondegenerate (it means that they are far from the possibility of BEC formation). Consequently, when densities of photons ${{n}_\textrm{ph}}\left( \mathbf{p} \right)$ and atoms $n_{{\alpha }_2}\left( \mathbf{p} \right)$ are below the transition temperatures ${{T}_\textrm{c}}$ and $T_\textrm{c}^{*}$ of the excited atoms and photons, correspondingly, the densities of distribution functions can take the form [analogous to \eqref{20}]:
\begin{equation} \begin{split}\begin{gathered}
\label{46}
{{n}_{{{\alpha }_{2}}}}\left( \mathbf{p} \right)=\frac{{{g}_{{{\alpha }_{2}}}}}{{{\left( 2\pi \hbar \right)}^{3}}}{{\left[ \exp \left( \frac{{{\mathbf{p}}^{2}}}{2mT} \right)-1 \right]}^{-1}}
+n_{{{\alpha }_{2}}}^{0}\left( T \right)\delta \left( \mathbf{p} \right),
\qquad
T\leqslant {{T}_\textrm{c}}\,,
\\
{{n}_\textrm{ph}}\left( \mathbf{p} \right)=\frac{{{g}_\textrm{ph}}}{{{\left( 2\pi \hbar \right)}^{3}}}{{\left[ \exp \left( \frac{{{\mathbf{p}}^{2}}}{2{{m}^{*}}T} \right)-1 \right]}^{-1}}
+n_\textrm{ph}^{0}\left( T \right)\delta \left( \mathbf{p} \right),
\qquad
 T\leqslant T_\textrm{c}^{*}\,.
\end{gathered} \end{split}\end{equation}
Here, $n_{{{\alpha }_{2}}}^{0}\left( T \right)$ is excited atoms BEC density and $n_\textrm{ph}^{0}\left( T \right)$ is photonic BEC density. The condition for excited atomic gas to be considered as nondegenerated one is formulated analogous to \eqref{14}, and it is given by the following:
\begin{equation} {
\label{47}
\exp \left( \frac{{{\varepsilon }_{{{\alpha }_{1}}}}-{{\mu }_{1}}}{T} \right)\gg 1.
} \end{equation}
Equations \eqref{46} were obtained in correspondence with the expressions below
\begin{equation} {
\label{48}
{{\mu }^{*}}\big| _{T\leqslant T_\textrm{c}^{*}} =\hbar {{\omega }_{0}}\,, \qquad {{\mu }_{2}}\big| _{T\leqslant {{T}_\textrm{c}}} ={{\varepsilon }_{2}}\,,
} \end{equation}
that are required for BEC of the system components to appear [see in this regard \cite{24} and \eqref{16}]. Formulae \eqref{46}, \eqref{47} allow us to rewrite \eqref{12} as follows:
\begin{equation} \begin{split}\begin{gathered}
\label{49}
n=n_{{{\alpha }_{2}}}^{0}\left( T \right)+{{g}_{{{\alpha }_2}}}\zeta \left( {3}/{2} \right){{\left( \frac{mT}{2\pi {{\hbar }^{2}}} \right)}^{3/2}}
+{{g}_{{{\alpha }_{1}}}}{{\left( \frac{mT}{2\pi {{\hbar }^{2}}} \right)}^{3/2}}\exp \left( \frac{\Delta }{T} \right),
\\
{{n}_\textrm{ph}}=\zeta \left( {3}/{2} \right){{\left( \frac{{{m}^{*}}T}{2\pi {{\hbar }^{2}}} \right)}^{3/2}}\left[ {{g}_\textrm{ph}}+{{g}_{{{\alpha }_{2}}}}{{\left( \frac{m}{{{m}^{*}}} \right)}^{3/2}} \right]
+
n_\textrm{ph}^{0}\left( T \right)+n_{{{\alpha }_{2}}}^{0}\left( T \right),
\end{gathered}
\end{split} \end{equation}
where $\Delta $ is still defined by \eqref{19}. We emphasize that in the case currently being studied, this value should be less than zero $\Delta <0$ to satisfy the condition \eqref{47} which makes it possible to regard the nonexcited atomic gas as nondegenerated one. Equations \eqref{49} give us condensate densities
$n_{{{\alpha }_{2}}}^{0}\left( T \right)$ and $n_\textrm{ph}^{0}\left( T \right)$ at $T\leqslant {{T}_\textrm{c}}$ and at $T\leqslant T_\textrm{c}^{*}$, correspondingly:
\begin{equation} \begin{split}\begin{gathered}
\label{50}
n_{{{\alpha }_{2}}}^{0}\left( T \right)=n-{{\left( \frac{mT}{2\pi {{\hbar }^{2}}} \right)}^{3/2}} {{g}_{{{\alpha }_{1}}}}\exp \left( -\frac{\left| \Delta \right|}{T} \right)
-
{{\left( \frac{mT}{2\pi {{\hbar }^{2}}} \right)}^{3/2}}{{g}_{{{\alpha }_{2}}}}\zeta \left( {3}/{2} \right) ,
\\
n_\textrm{ph}^{0}\left( T \right)={{n}_\textrm{ph}}-n+{{g}_{{{\alpha }_{1}}}}{{\left( \frac{mT}{2\pi {{\hbar }^{2}}} \right)}^{3/2}}\exp \left( -\frac{\left| \Delta \right|}{T} \right)
-{{g}_\textrm{ph}}\zeta \left( {3}/{2}\right){{\left( \frac{{{m}^{*}}T}{2\pi {{\hbar }^{2}}} \right)}^{3/2}}.
\end{gathered}
\end{split} \end{equation}
This can also provide us with transition temperatures ${{T}_\textrm{c}}$ and $T_\textrm{c}^{*}$, if we take into account that the densities of condensate become zero in transition points.
\begin{equation} \begin{split}\begin{gathered}
\label{51}
n_{{{\alpha }_{2}}}^{0}\left( {{T}_\textrm{c}} \right)=0 ,\qquad n_\textrm{ph}^{0}\left( T_\textrm{c}^{*} \right)=0 .
\end{gathered}
\end{split} \end{equation}
Let us remark here that since the photon condensate density should have a positive value (including the case when $T\to 0$), the second equation in \eqref{50} implies the following inequality:
\begin{equation}
\label{52}
{{n}_\textrm{ph}}>n.
\end{equation}
Later on in this article we shall get back to the question of circumstances when $n_\textrm{ph}^{0}\left( T \right)$ has a positive value in all the allowed temperatures range.

For definiteness it is assumed that $T_\textrm{c}^{*}<{{T}_\textrm{c}}$ (as we have done in the previous section). The expressions to define critical temperatures look as follows:
\begin{equation} \begin{split}\begin{gathered}
\label{53}
n=\!{{\left( \frac{m{{T}_\textrm{c}}}{2\pi {{\hbar }^{2}}} \right)}^{3/2}}\! \left[ {{g}_{{{\alpha }_1}}}\!\exp \left( -\frac{\left| \Delta \right|}{{{T}_\textrm{c}}} \right)+{{g}_{{{\alpha }_2}}}\zeta \left( {3}/{2} \right) \right],
\\
{{n}_\textrm{ph}}-n={{g}_\textrm{ph}}\zeta \left( {3}/{2}\right){{\left( \frac{{{m}^{*}}T_\textrm{c}^{*}}{2\pi {{\hbar }^{2}}} \right)}^{3/2}}
-{{g}_{{{\alpha }_{1}}}}{{\left( \frac{mT_\textrm{c}^{*}}{2\pi {{\hbar }^{2}}} \right)}^{3/2}}\exp \left( -\frac{\left| \Delta \right|}{T_\textrm{c}^{*}} \right).
\end{gathered}
\end{split} \end{equation}
When temperatures are supposed to be low
\begin{equation*}
\exp \left( -\frac{\left| \Delta \right|}{{{T}_\textrm{c}}} \right)\ll 1
\end{equation*}
one may obtain the result from the first equation in \eqref{53}
\begin{equation}  \begin{split}\begin{gathered}
\label{54}
{{T}_\textrm{c}}\approx \frac{2\pi {{\hbar }^{2}}}{m}{{\left[ \frac{n}{{{g}_{{{\alpha }_{2}}}}\zeta \left( {3}/{2}\; \right)} \right]}^{2/3}} .
\end{gathered}
\end{split} \end{equation}
It is easy to see that if we replace coefficient ${g_{{\alpha }_1}}$with ${g_{{\alpha }_2}}$, the expression \eqref{54} coincides with the similar one for a transition temperature \eqref{24} in the main approximation of the first order of smallness. It is clear that in the case being studied, the approximation tool adaptability criteria are defined by the relation \eqref{26} where one needs to replace ${g_{{\alpha }_1}}$ with ${g_{{\alpha }_2}}$, and $\Delta $ value needs to be replaced with its absolute value~$\left| \Delta \right|$.

Let us analyze the second equation in \eqref{53}; when using \eqref{54}, it can be rearranged into the form:
\begin{equation}   \begin{split}\begin{gathered}
\label{55}
\frac{{{n}_\textrm{ph}}-n}{n}={{\left( \frac{T_\textrm{c}^{*}}{{{T}_\textrm{c}}} \right)}^{3/2}}\left[ \frac{{{g}_\textrm{ph}}}{{{g}_{{{\alpha }_{2}}}}}{{\left( \frac{{{m}^{*}}}{m} \right)}^{3/2}}
-\frac{{{g}_{{{\alpha }_{1}}}}}{{{g}_{{{\alpha }_{2}}}}\zeta \left( {3}/{2}\right)}\exp \left( -\frac{\left| \Delta \right|}{{{T}_\textrm{c}}}\frac{{{T}_\textrm{c}}}{T_\textrm{c}^{*}} \right) \right] , \qquad T_\textrm{c}^{*}<{{T}_\textrm{c}} \,.
\end{gathered}
\end{split} \end{equation}
For simplicity in further calculations we shall assume that $T_\textrm{c}^{*}<{{T}_\textrm{c}}$ and the temperatures ${{T}_\textrm{c}}$ and $T_\textrm{c}^{*}$ are of the same order of magnitude ${T_\textrm{c}^{*}}/{T_\textrm{c}}\sim 1$; equation \eqref{55} takes the form:
\begin{equation*} \begin{split}\begin{gathered}
\frac{{n_\textrm{ph}}-n}{n}\approx \frac{{g_\textrm{ph}}}{{g_{{{\alpha }_2}}}}{{\left( \frac{{{m}^{*}}}{m} \right)}^{3/2}}-\frac{{{g}_{{{\alpha }_{1}}}}}{{{g}_{{{\alpha }_{2}}}}\zeta \left( {3}/{2} \right)}\exp \left( -\frac{\left| \Delta \right|}{T_\textrm{c}^{*}} \right) ,
\end{gathered}
\end{split} \end{equation*}
from which we get the following formula:
\begin{equation}  \begin{split}\begin{gathered}
\label{56}
\!\!\exp \left( -\frac{\left| \Delta \right|}{T_\textrm{c}^{*}} \right)\!\approx \! \zeta \left( {3}/{2} \right) \! \left[ \frac{{g_\textrm{ph}}}{{g_{\alpha _1}}}{{\left( \frac{{{m}^{*}}}{m} \right)}^{3/2}}-\frac{{{g}_{{{\alpha }_2}}}}{{{g}_{{{\alpha }_1}}}}\frac{{{n}_\textrm{ph}}-n}{n} \right] ,
\end{gathered}
\end{split} \end{equation}
this evidently implies that inequalities are valid:
\begin{equation}    \begin{split}\begin{gathered}
\label{57}
{{\left( \frac{{{m}^{*}}}{m} \right)}^{3/2}}>\frac{{{g}_{{{\alpha }_{2}}}}}{{{g}_\textrm{ph}}}\frac{{{n}_\textrm{ph}}-n}{n}\,,
\\
\qquad \zeta \left( {3}/{2} \right)\left[ \frac{{{g}_\textrm{ph}}}{{{g}_{{{\alpha }_{1}}}}}{{\left( \frac{{{m}^{*}}}{m} \right)}^{3/2}}-\frac{{{g}_{{{\alpha }_{2}}}}}{{{g}_{{{\alpha }_{1}}}}}\frac{{{n}_\textrm{ph}}-n}{n} \right]\ll 1.
\end{gathered}
\end{split} \end{equation}
The first one in \eqref{57} is trivial; the second one results from inequality $\exp \left( -{\left| \Delta \right|}/{{{T}_\textrm{c}}} \right)\ll 1$ when considering ${T_\textrm{c}^{*}}/{{{T}_\textrm{c}}}\sim 1$. In addition, the expression ${{n}_\textrm{ph}}\sim n$ in order to regard atoms as two level atoms needs to be valid. As it can be seen from \eqref{57}, \eqref{52} the value of $({{{n}_\textrm{ph}}-n})/{n}$ has the upper limit:
\begin{equation}
\label{58}
\frac{{{n}_\textrm{ph}}-n}{n}\lesssim\frac{{{g}_\textrm{ph}}}{{{g}_{{{\alpha}_{2}}}}}{{\left(\frac{{{m}^{*}}}{m}\right)}^{3/2}}.
 \end{equation}
Such limitation means that the value of $({{{n}_\textrm{ph}}-n})/{n}$ is negligibly small. Actually, as we mentioned above, ${{{m}^{*}}}/{m}$ ratio is very small [see formulae \eqref{28}--\eqref{32}]; for lithium atom this value is $\left( {{{m}^{*}}}/{m} \right)\sim {{10}^{-10}}$, and consequently for this case we have $({{{n}_\textrm{ph}}-n})/{n}\lesssim{{10}^{-15}}$.

\begin{figure}
\centerline{\includegraphics[width=0.7\textwidth]{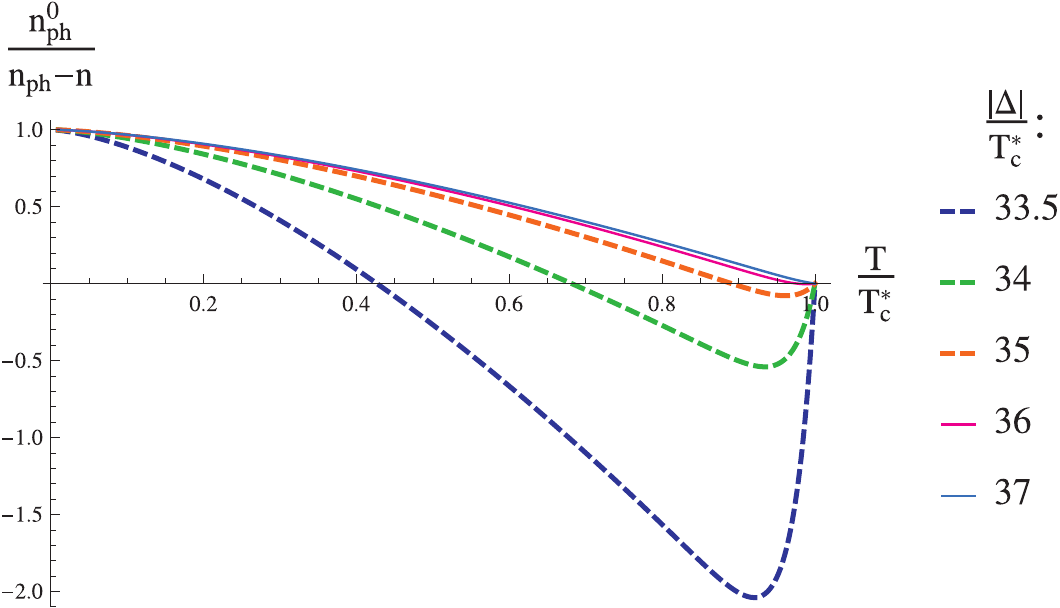}}
\caption{(Color online) Illustration of nonmonotonous dependence of condensate density $n_\textrm{ph}^{0}\left( T \right)$ on the temperature for different $\Delta$ values.
}
\label{third}
\end{figure}

Equation \eqref{56} allows us to find the analytical expression for the transition temperature $T_\textrm{c}^{*}$ that should be of the same order of magnitude as ${{T}_\textrm{c}}$
\begin{equation} {
\label{59}
T_\textrm{c}^{*}\approx \frac{- \left| \Delta \right|}{\ln {{\left\{ \zeta \left( {3}/{2} \right)\left[ \frac{{{g}_\textrm{ph}}}{{{g}_{{{\alpha }_{1}}}}}{{\left( \frac{{{m}^{*}}}{m} \right)}^{3/2}}-\frac{{{g}_{{{\alpha }_{2}}}}}{{{g}_{{{\alpha }_{1}}}}}\frac{{{n}_\textrm{ph}}-n}{n} \right] \right\}}^{}}}, \qquad \frac{T_\textrm{c}^{*}}{{{T}_\textrm{c}}}\lesssim1.
} \end{equation}
Let us get back to the question about positivity of photon condensate density [see \eqref{50}] at any temperature below the condensation temperature. It is easy to analyze that condition \eqref{52} does not provide this value positivity at any temperature, because $n_\textrm{ph}^{0}\left( T \right)$ is a nonmonotonous function and, consequently, not all values of ${\left| \Delta \right|}/{T_\textrm{c}^{*}}\gg 1$ ensure the photonic condensate density to be positive. The evidence of that can be seen in figure~\ref{third}: here, $n_\textrm{ph}^{0}\left( T \right)$ dependences were plotted according to expressions \eqref{50}, \eqref{53} for some values of ${\left| \Delta \right|}/{T_\textrm{c}^{*}}$ when it was still supposed that ${\left| \Delta \right|}/{T_\textrm{c}^{*}}\gg 1$ (the cases when density $n_\textrm{ph}^{0}\left( T \right)$ is negative in some temperature range were drawn with dashed lines). From figure~\ref{third} one may see that $n_\textrm{ph}^{0}\left( T \right)$  is positively defined if ${\left| \Delta  \right|}/{T_\textrm{c}^{*}}\gtrsim36$.
According to the ``hint'' given in figure~\ref{third} we can find a more precise limitation for the studied system parameters which provides positivity of $n_\textrm{ph}^{0}\left( T \right)$ along the whole range of the allowed temperatures. To do so, let us suppose that when $T=T_\textrm{c}^{*}$, the derivative ${\partial n_\textrm{ph}^{0}\left( T \right)}/{\partial T}$ is negative
\begin{equation} {
\label{60}
{{\left. \frac{\partial }{\partial T}n_\textrm{ph}^{0}\left( T \right) \right|}_{T=T_\textrm{c}^{*}}}<0 .
} \end{equation}
As we can see from figure~\ref{third} such a claim certainly provides us with $n_\textrm{ph}^{0}\left( T \right)$ value monotonous increase when temperature is lowering within $0<T\leqslant T_\textrm{c}^{*}$ range. By calculating the derivative ${{ \big[ {\partial n_\textrm{ph}^{0}\left( T \right)}/{\partial T} \big] \big|}_{T=T_\textrm{c}^{*}}}$ from \eqref{50} and using the second equation in \eqref{53} one may rewrite expression \eqref{60} as follows:
\begin{equation*}
\frac{2}{3}\frac{\left| \Delta \right|}{T_\textrm{c}^{*}}\!\exp \left(\! -\frac{\left| \Delta \right|}{T_\textrm{c}^{*}} \right)+\exp \left( -\frac{\left| \Delta \right|}{T_\textrm{c}^{*}} \right)<\frac{{{g}_\textrm{ph}}}{{{g}_{{{\alpha }_{1}}}}}\zeta \left( {3}/{2}\right){{\left( \frac{{{m}^{*}}}{m}\right)}^{3/2}},
\end{equation*}
and when ${\left| \Delta \right|}/{T_\textrm{c}^{*}}\gg 1$ we get:
\begin{equation} {
\label{61}
\frac{\left| \Delta \right|}{T_\textrm{c}^{*}}\exp \left( -\frac{\left| \Delta \right|}{T_\textrm{c}^{*}} \right)<\frac{3}{2}\frac{{{g}_\textrm{ph}}}{{{g}_{{{\alpha }_1}}}}\zeta \left( {3}/{2} \right){{\left( \frac{{{m}^{*}}}{m} \right)}^{3/2}} .
} \end{equation}
This is the inequality \eqref{61} that defines [in addition to \eqref{57}, \eqref{58}] another coexistence condition of photonic BEC and BEC of excited atoms when atoms in the ground state are nondegenerated and all components are in thermodynamic equilibrium. Besides this, relation \eqref{61} allows us to find an approximated formula of photonic condensate density in the system studied:
\begin{equation} {
\label{62}
n_\textrm{ph}^{0}\left( T \right)\approx \left( {{n}_\textrm{ph}}-n \right)\left[ 1-{{\left( \frac{T}{T_\textrm{c}^{*}} \right)}^{3/2}} \right] .
} \end{equation}
The density of excited atoms BEC according to \eqref{50}, \eqref{54} looks as follows:
\begin{equation} {
\label{63}
n_{{{\alpha }_{2}}}^{0}\left( T \right)\approx n\left[ 1-{{\left( \frac{T}{{{T}_\textrm{c}}} \right)}^{3/2}} \right] ,
} \end{equation}
and density of atoms ${{n}_{{{\alpha }_{1}}}}\left( T \right)$ in the ground state will be exponentially decreased by lowering the temperature, and density takes the following form [see a similar approach in \eqref{49}]:
\begin{equation} {
\label{64}
{{n}_{{{\alpha }_{1}}}}\left( T \right)\approx n\frac{{{g}_{{{\alpha }_{1}}}}}{{{g}_{{{\alpha }_{2}}}}\zeta \left( {3}/{2}\; \right)}{{\left( \frac{T}{{{T}_\textrm{c}}} \right)}^{3/2}}\exp \left( -\frac{\left| \Delta \right|}{T} \right) .
} \end{equation}
Formulae \eqref{62}--\eqref{64} generally solve the problem announced in this section for coexistence of BEC of excited atoms and photonic BEC when ground state atoms are nondegenerated. However, we have to mention that due to limitation \eqref{58} photons BEC density in this case is negligibly small compared with the atoms in the excited state BEC density. For this reason~--- considering the problem stated in this article (BEC coexistence conditions study)~--- the case appears to be of little interest to us. It can have some interest for another reason: as it is easy to see from \eqref{62}--\eqref{64} within temperatures range
\begin{equation}
\label{65}
\frac{T}{{{T}_\textrm{c}}}\ll 1,\qquad \frac{T_\textrm{c}^{*}}{{{T}_\textrm{c}}}<1 , \qquad \frac{{{T}_\textrm{c}}}{\left| \Delta \right|}\ll 1,
\end{equation}
almost all photons in the system are basically absorbed by atoms in the ground state within the possibility of their transfer to the excited state; almost all excited atoms form atomic BEC in the system and that is why the number of atoms in the ground state becomes exponentially small depending on the temperature. In other words, when the temperature of the studied system  lowers, the population inversion of atomic levels takes place with the formation of BEC of photons and excited atoms. Moreover, such pumping occurs in an abrupt type mode of condensation according to \eqref{64} taking into account expression \eqref{65}. Note, that such an abrupt pumping looks rather unexpected: the number of excited atoms increases when temperature decreases. This situation can be also treated as ``stopped light'' (actually electromagnetic waves are meant here), because group velocity of photons in BEC is equal to zero [see \eqref{45}]. Nevertheless, in this section such a statement is less significant than it was within the previous section: now we can have a few photons in the system and consequently, a few photons can be seen in the condensate. The described situation is likely to be considered as the one having relevance to the storage of light in atomic vapor at ultralow temperatures \cite{10,11}.

As the next step to complete the overall picture of our research, we shall study the possibility of simultaneous coexistence of three BEC at the same temperature of system components (photonic and two atomic components).

\section{On the possibility of BEC co-existence in all system components}
The equations \eqref{12}, \eqref{13} can be considered as the initial ones to study this case. To obtain Bose Einstein condensates in three subsystems, the following is needed see \cite{24} and \eqref{16}, \eqref{48}:
\begin{equation}
\label{66}
{{\mu }^{*}\big| _{T\leqslant T_\textrm{c}^{*}} =\hbar {{\omega }_{0}}\,, \qquad {{\mu }_{1}}\big| _{T\leqslant {{T}_\textrm{c}}} ={{\varepsilon }_{1}}\,, \qquad {{\mu }_{2}}\big| _{T\leqslant {{T}_\textrm{c}}} ={{\varepsilon }_{2}}}\,.
\end{equation}
Equations \eqref{66} change the third one in \eqref{9} as follows:
\begin{equation*}
{{\varepsilon }_{{{\alpha }_2}}}={{\varepsilon }_{{\alpha }_1}}+\hbar {{\omega }_0}
\end{equation*}
or
\begin{equation}
\label{67}
\Delta =0,
\end{equation}
where $\Delta $ is still defined by \eqref{19}. As it follows from two previous sections, if \eqref{67} is valid, then two atomic components can form BEC simultaneously at the same temperature ${{T}_\textrm{c}}$. We showed that if the inequality $\Delta >0$ fulfills, it makes possible to achieve BEC in photonic and ground state atomic subsystem, whilst the validity of $\Delta <0$ provides coexistence of photons and excited atoms in BEC. From this particular fact we come to a conclusion that BEC formation of two atomic gas components is only possible at the same temperature for both.

When temperature is below ${{T}_\textrm{c}}$ and $T_\textrm{c}^{*}$the latter statement gives us the following distribution functions densities of atoms ${{n}_{{{\alpha }_1}}}\left( \mathbf{p} \right)$, ${{n}_{{{\alpha }_2}}}\left( \mathbf{p} \right)$ and photons ${{n}_\textrm{ph}}\left( \mathbf{p} \right)$ [see \eqref{10}, \eqref{11}]; these functions can take the form [see also \eqref{20}, \eqref{49}]:
\begin{equation} \begin{split}\begin{gathered}\begin{aligned}
\label{68}
&{{n}_{{{\alpha }_1}}}\left( \mathbf{p} \right)=n_{{{\alpha }_1}}^{0}\left( T \right)\delta \left( \mathbf{p} \right)+\frac{{{g}_{{{\alpha }_{1}}}}}{{{\left( 2\pi \hbar \right)}^{3}}}{{\left[ \exp \left( \frac{{{\mathbf{p}}^{2}}}{2mT} \right)-1 \right]}^{-1}},
\qquad T\leqslant {{T}_\textrm{c}} ,
\\
&{{n}_{{{\alpha }_2}}}\left( \mathbf{p} \right)=n_{{{\alpha }_2}}^{0}\left( T \right)\delta \left( \mathbf{p} \right)+\frac{{{g}_{{{\alpha }_{2}}}}}{{{\left( 2\pi \hbar \right)}^{3}}}{{\left[ \exp \left( \frac{{{\mathbf{p}}^{2}}}{2mT} \right)-1 \right]}^{-1}},
\\
&{{n}_\textrm{ph}}\left( \mathbf{p} \right)=n_\textrm{ph}^{0}\left( T \right)\delta \left( \mathbf{p} \right)+\frac{{{g}_\textrm{ph}}}{{{\left( 2\pi \hbar \right)}^{3}}}{{\left[ \exp \left( \frac{{{\mathbf{p}}^{2}}}{2{{m}^{*}}T} \right)-1 \right]}^{-1}} ,
\qquad  T\leqslant T_\textrm{c}^{*},
 \end{aligned}\end{gathered}\end{split}\end{equation}
where $n_{{{\alpha }_{1}}}^{0}\left( T \right)$, $n_{{{\alpha }_{2}}}^{0}\left( T \right)$, $n_\textrm{ph}^{0}\left( T \right)$ are the densities of condensates of atomic components and free photons, correspondingly. By inserting \eqref{68} into \eqref{13} one may obtain the next system of equations at $T<{{T}_\textrm{c}}$, $T<T_\textrm{c}^{*}$ temperatures:
\begin{equation} \begin{split}\begin{gathered}
\label{69}
n=n_{{{\alpha }_{1}}}^{0}+n_{{{\alpha }_{2}}}^{0}+\zeta \left( {3}/{2}\right)\left( {{g}_{{{\alpha }_{1}}}}+{{g}_{{{\alpha }_{2}}}} \right){{\left( \frac{mT}{2\pi {{\hbar }^{2}}} \right)}^{3/2}} ,
\\
{{n}_\textrm{ph}}=n_\textrm{ph}^{0}+n_{{{\alpha }_{2}}}^{0}
+\zeta \left( {3}/{2}\right){{\left( \frac{{{m}^{*}}T}{2\pi {{\hbar }^{2}}} \right)}^{3/2}}\left[ {{g}_\textrm{ph}}+{{g}_{{{\alpha }_{2}}}}{{\left( \frac{m}{{{m}^{*}}} \right)}^{3/2}} \right] .
  \end{gathered}\end{split}\end{equation}
This system of equations is not complete: it includes two equations for the three unknown [$n_{{{\alpha }_1}}^{0}\left( T \right)$, $n_{{{\alpha }_{2}}}^{0}\left( T \right)$ and $n_\textrm{ph}^{0}\left( T \right)$]. We can easily add the third equation to the system \eqref{69} if we notice that the first equation in \eqref{13} and the definitions \eqref{68} when ${{\mu }_{1}}\big| _{T\leqslant {{T}_\textrm{c}}} ={{\varepsilon }_{1}}$ and ${{\mu }_{2}}\big| _{T\leqslant {{T}_\textrm{c}}} ={{\varepsilon }_2}$ give us the following.
Performing easy calculations using \eqref{68} the latter equation can be rearranged into the following form:
\begin{equation} {
\label{70}
{{g}_{{{\alpha }_2}}}n_{{{\alpha }_1}}^{0}={g_{{{\alpha }_1}}}n_{{{\alpha }_2}}^{0} \,.
} \end{equation}
This expression can be regarded as the one we lacked to make \eqref{69} a closed system of equations. The solutions of \eqref{69}, \eqref{70} are given by:
\begin{equation} \begin{split}\begin{gathered}
\label{71}
n_\textrm{ph}^{0}\left( T \right)={n_\textrm{ph}}-\frac{n{g_{{{\alpha }_2}}}}{{g_{{{\alpha }_1}}}+{g_{{{\alpha }_2}}}}-\zeta \left( {3}/{2} \right){{g}_\textrm{ph}}{{\left( \frac{{{m}^{*}}T}{2\pi {{\hbar }^2}} \right)}^{3/2}} ,
\\
n_{{{\alpha }_1}}^{0}\left( T \right)={g_{{{\alpha }_{1}}}}\left[ \frac{n}{{{g}_{{{\alpha }_{1}}}}+{g_{{{\alpha }_2}}}}-\zeta \left( {3}/{2} \right){{\left( \frac{mT}{2\pi {{\hbar }^{2}}} \right)}^{3/2}} \right] ,
\\
n_{{{\alpha }_2}}^{0}\left( T \right)={g_{{{\alpha }_2}}}\left[ \frac{n}{{g_{{{\alpha }_1}}}+{g_{{{\alpha }_2}}}}-\zeta \left( {3}/{2} \right){{\left( \frac{mT}{2\pi {{\hbar }^{2}}} \right)}^{3/2}} \right] .
 \end{gathered}\end{split}\end{equation}
One may obtain the transition temperatures of system components from the previous formulae by taking into account that the Bose condensates disappear in the transition point:
\begin{equation} {
\label{72}
n_{{{\alpha }_1}}^{0}\left( {{T}_\textrm{c}} \right)=0 , \qquad n_{{{\alpha }_{2}}}^{0}\left( {{T}_\textrm{c}} \right)=0 , \qquad n_\textrm{ph}^{0}\left( T_\textrm{c}^{*} \right)=0 .
} \end{equation}
The condition $\Delta =0$ [see \eqref{67}] leads to the similar temperature ${{T}_\textrm{c}}$ of transition to BEC states for both atomic components as
\begin{equation} {
\label{73}
{{T}_\textrm{c}}=\frac{2\pi {{\hbar }^{2}}}{m}{{\left[ \frac{n}{\zeta \left( {3}/{2}\; \right)\left( {{g}_{{{\alpha }_{1}}}}+{{g}_{{{\alpha }_{2}}}} \right)} \right]}^{{2}/{3}\;}} .
} \end{equation}
and it is easy to see that this condition follows from the last two equations in \eqref{48}. In the studied system, photon condensation temperature can be calculated using \eqref{72} and taking into account \eqref{71}:
\begin{equation} {
\label{74}
T_\textrm{c}^{*}=\frac{2\pi {{\hbar }^{2}}}{{{m}^{*}}}{{\left[ \frac{n_\textrm{ph}^\textrm{eff}}{\zeta \left( {3}/{2} \right){{g}_\textrm{ph}}} \right]}^{{2}/{3}}} ,
} \end{equation}
where $n_\textrm{ph}^\textrm{eff}$ is some effective density of photons and it is defined by a formula
\begin{equation} {
\label{75}
n_\textrm{ph}^\textrm{eff}\equiv {{n}_\textrm{ph}}-n\frac{{g_{{{\alpha }_2}}}}{{{g}_{{{\alpha }_1}}}+{g_{{{\alpha }_2}}}} \,.
} \end{equation}
Since photon condensation density should be positive at any temperature including $T\to 0$ from \eqref{71} we get density limit of photons and atoms in the system to make the coexistence of BEC in all three system components possible:
\begin{equation} {
\label{76}
{{n}_\textrm{ph}}>n\frac{{{g}_{{{\alpha }_{2}}}}}{{{g}_{{{\alpha }_{1}}}}+{{g}_{{{\alpha }_{2}}}}} \,.
} \end{equation}
Besides, if condensation temperatures are supposed to satisfy the inequality $T_\textrm{c}^{*}<{{T}_\textrm{c}}$ but to have the same order of magnitude $T_\textrm{c}^{*}\sim {{T}_\textrm{c}}$, one can get a more precise limitation of $n_\textrm{ph}^\textrm{eff}$ rather than \eqref{76} [see \eqref{75}]. From \eqref{73}, \eqref{64} we can obtain
\begin{equation} {
\label{77}
\frac{T_\textrm{c}^{*}}{{{T}_\textrm{c}}}=\frac{m}{{{m}^{*}}}{{\left[ \frac{n_\textrm{ph}^\textrm{eff}}{n}\left( \frac{{{g}_{{{\alpha }_{1}}}}+{{g}_{{{\alpha }_{2}}}}}{{{g}_\textrm{ph}}} \right) \right]}^{{2}/{3} }} .
} \end{equation}
As we have mentioned several times, the masses ratio ${{{m}^{*}}}/{m}$ is extremely small [for lithium $\left( {{{m}^{*}}}/{m} \right)\sim {{10}^{-10}}$, see above] and, consequently, ${m}/{{{m}^{*}}}$ is rather high; for that reason, to satisfy $T_\textrm{c}^{*}\sim {{T}_\textrm{c}}$ one needs the densities ratio ${n_\textrm{ph}^\textrm{eff}}/{n}$ in \eqref{77} to be small [like in \eqref{36}, \eqref{37}]:
\begin{equation} {
\label{78}
\frac{n_\textrm{ph}^\textrm{eff}}{n}\sim {{10}^{-15}} .
} \end{equation}
In terms of the critical temperatures ${{T}_\textrm{c}}$, $T_\textrm{c}^{*}$ [see \eqref{73}, \eqref{74}], the expressions \eqref{71} for condensate densities of all three system components may be rewritten in a more common way (see for example \cite{24})
\begin{equation} \begin{split}\begin{gathered}
\label{79}
n_\textrm{ph}^{0}={{n}_\textrm{eff}}\left[ 1-{{\left( \frac{T}{T_\textrm{c}^{*}} \right)}^{3/2}} \right] ,
\\
n_{{{\alpha }_{1}}}^{0}=\frac{{{g}_{{{\alpha }_{1}}}}}{{{g}_{{{\alpha }_{1}}}}+{{g}_{{{\alpha }_{2}}}}}n\left[ 1-{{\left( \frac{T}{{{T}_\textrm{c}}} \right)}^{{3}/{2} }} \right],
\\
n_{{{\alpha }_{2}}}^{0}=\frac{{{g}_{{{\alpha }_{2}}}}}{{{g}_{{{\alpha }_{1}}}}+{{g}_{{{\alpha }_{2}}}}}n\left[ 1-{{\left( \frac{T}{{{T}_\textrm{c}}} \right)}^{{3}/{2} }} \right] .
\end{gathered}\end{split} \end{equation}
From \eqref{79}, \eqref{78} we can conclude that photons condensate density in the system under the conditions being studied is negligibly small compared with the atomic condensate density. This is the reason why in this section the studied case is not of great interest in the context of atomic and photonic BEC coexistence.

\section{Conclusion}

By this means, we have studied all three possible variants when photonic Bose condensate coexists in thermodynamic equilibrium with Bose condensate of ideal two-level atomic gas. That is to say, it was supposed that photonic BEC can always be created in the system, although the critical temperature of photons was considered to be below the atomic one. In the context of atomic Bose-Einstein condensation, three cases were assumed to be possible for implementation:
\begin{enumerate}
 \item BEC can be formed by ground state atoms with nondegenerate gas component of excited atoms.
\item  BEC can be formed by excited atoms with nondegenerate gas component of ground state atoms.
\item Both atomic components form BEC.
\end{enumerate}

For all these situations we have found critical temperatures, densities of condensates for atomic and photon components and conditions of their coexistence. The first case (when condensates of ground state atoms and photons coexist) was shown to be the most effective from experimental implementation point of view, because an abrupt photon condensation when temperature lowers was predicted for this case. This situation, in the authors opinion, is the closest to the one which can be treated as stopped light in BEC. Also, it was shown that other cases may also concern the storage of light in atomic vapors at ultralow temperatures.

Some features of our system model need to be improved. For example, in this article we have used the approach common enough in theoretical physics, optics and photonics, i.e., atoms of the gas were treated as the two-level ones. Besides, from the standpoint of the current article such an assumption is not crucial: it only  greatly simplifies the calculations and even allows one to get some analytical results of calculations. Equations of such type as \eqref{9} or \eqref{12} may be formulated for an arbitrary large number of components, but in this case a complicated problem appears, i.e., to trace such numerical calculations for these equations.

There are some interactions of atoms and some scattering processes that can influence the coexistence conditions of the studied condensates. However, it is a rather challenging task to take such interactions into account, which is, in our opinion, beyond the scope of this paper. Moreover, such a task must be based on microscopic approach that may be provided by quantum electrodynamics (at low temperatures). In our opinion, some promising approach was developed and presented in \cite{25}; we are currently studying such an issue.

We supposed that the thermodynamic equilibrium certainly exists in the system when  we formulated and solved the problem, but such a premise is not obvious and is something to be discussed.  For example, some difficulties concerning the thermalization process  at low temperatures were pointed out in \cite{27}: here,  nonequilibrium model of photon condensation in a dye filled optical microcavity was studied. Some similar peculiarities may exist when thermalization happens in the  system treated by us. However, to answer this question we need to  take  interactions between system components into account and to solve nonequilibrium problem of condensation of photons. We think that such a question is far beyond  the scope of this paper.

It should be noted that one may also set a problem of forming photonic BEC in thermodynamic equilibrium with ideal Fermi gas. Let us recall that equations \eqref{9} were written for both Fermi and Bose gases being in thermodynamic equilibrium with photons. These are the initial equations that were provided by formulae \eqref{9} in the case of photon Bose condensation in thermodynamic equilibrium with degenerated Fermi gas. Preliminary estimates show that in the latter case photon condensate densities are negligibly small compared to the atomic ones; also, photon transition temperatures may be much lower than common temperatures of Fermi gas degeneration. However, such estimates were made by the authors in the area of physical
parameters that were suitable for analytical calculations. This fact makes such a study a separate task, which claims more detailed research using special numerical methods.

\ukrainianpart
 \title{Співіснування фотонного й атомарного бозе-ейнштейнівських конденсатів \\в ідеальних атомарних газах}
 \author{Н. Бойченко\refaddr{label1},
 Ю. Слюсаренко\refaddr{label1, label2}}
 \addresses{
 \addr{label1} Інститут теоретичної фізики ім.~О.І.~Ахієзера ННЦ ХФТІ, вул. Академічна, 1, 61108 Харків, Україна
 \addr{label2} Харківський національний університет ім.~В.Н.~Каразіна, пл.~Свободи,~4, 61077 Харків, Україна
 }

 \makeukrtitle

 \begin{abstract}
 Дослiджено умови утворення конденсату Бозе-Ейнштейна для фотонів, які знаходяться у стані термодинамічної рівноваги з ідеальним газом дворівневих бозе-атомів нижче температури виродження. Сформульовано рівняння, що описують умови термодинамічної рівноваги в системі. Отримано аналітичні вирази для критичних температур та густин конденсатів у фотонних і атомарних підсистемах. Визначено умови спiвiснування бозе-конденсатiв в атомарних i фотонних компонентах. Передбачено можливість ``різкого'' режиму конденсації фотонів за присутності бозе-конденсату атомів в основному стані: показано, що навіть мале зниження температури може привести до значного збільшенню фотонів у конденсаті. Цей випадок пропонується розглядати в якості простої моделі ситуації, відомої як ``зупинене світло'' у холодному атомарному газі. Вказано на можливість інверсної заселеності атомних рівнів шляхом зниження температури. Дане явище виглядає перспективним з точки зору накопичення світла в атомарних газах за наднизьких температур.
 \keywords ідеальний газ, термодинамічна рівновага, бозе-ейнштейнівський конденсат фотонів, співіснування бозе-ейнштейнівських конденсатів
 \end{abstract}

\end{document}